\documentclass[preprint]{revtex4}
\usepackage{graphicx}
\usepackage{subfigure}
\usepackage{multirow}
\usepackage{hyperref}
\usepackage{amsmath,bm,amssymb,amsfonts}
\usepackage{color}
\newcommand{\dalm}{\kern1pt\vbox{\hrule height 0.9pt\hbox{\vrule width 0.9pt
			\hskip 2.5pt\vbox{\vskip 5.5pt}\hskip 3pt\vrule width 0.3pt}\hrule height 0.3pt}
	\kern1pt}
\begin{document}
	\title{{\bf
			Strange Quark Stars and Spontaneous Scalarization}}
	
	\author{{\bf Mina Mehraeen Nazdik $^{1,2}$ } and {\bf Zeinab Rezaei $^{1,2}$} \footnote{Corresponding author. E-mail:
				zrezaei@shirazu.ac.ir}}
	\affiliation{ $^{1}$Department of Physics, College of Science, Shiraz University, Shiraz 71454, Iran.\\
		$^{2}$Biruni Observatory, College of Science, Shiraz
University, Shiraz 71454, Iran.\\}
	
	%%%%%%%%%%%%%%%%%%%%%%%%%%%%%%%%%%%%%%%%%%%%%%%%%%%%%%%%%%%%%%%%%%%%%%%%%
	
	\begin{abstract}
Strange quark matter, as one of the most extreme dense states in astrophysics, can be explored through relativistic compact objects. Hypothetical relativistic stars composed of dense strange quark matter are characterized by the equation of state of that matter. The spontaneous scalarization of strange quark stars is influenced by the equation of state of strange quark matter. Here, we investigate how the equation of state of strange quark matter influences the spontaneous scalarization of these stars. Considering Normal, color-flavor-locked (CFL), and CFLm quark matter in the bag models and applying the scalar-tensor gravity, we calculate the structure of scalarized strange quark stars. We investigate how the effective bag constant, perturbative QCD parameter, CFL pairing gap, and strange quark mass control the mass-radius relation, the central scalar field, the scalarization thresholds, and the scalar charge. Our results demonstrate that strange quark stars in scalar-tensor gravity are viable, testable objects, and they establish a complete framework for constraining both the gravitational theory and the properties of cold quark matter with current and future multimessenger observations.

	\end{abstract}
	
	\maketitle
	%%%%%%%%%%%%%%%%%%%%%%%%%%%%%%%%%%%%%%%%%%%%%%%%%%%%%%%
\section{Introduction} \label{sec:intro}

If strange quark matter is energetically favored as the ground state of strongly interacting matter at high density and low temperature, compact stars composed of such matter can exist as strange quark stars (SQSs).
The (2+1)-flavor Nambu-Jona-Lasinio model admits stable SQS configurations whose macroscopic properties can be compatible with observational constraints from PSR J0740+6620 and the merger event GW170817 \cite{arXiv:1912.05093}.
Color-flavor-locked (CFL) SQS models have been shown to reproduce the macroscopic properties inferred from compact-object observations, including masses, radii, and tidal deformabilities where available, for systems such as GW170817, GW190425 \cite{arXiv:2304.12209}, and GW190814 \cite{arXiv:2010.11020}. These results indicate that CFL SQS models are compatible with current observational constraints, but they do not uniquely establish that these objects are SQSs.
Furthermore, the secondary component of GW190814 has been proposed as a possible up-down quark star candidate \cite{arXiv:2009.00942} or a dark matter admixed SQS \cite{arXiv:2209.09021,arXiv:2402.14262}.
Dark matter admixed SQS models have been shown to reproduce compact-star properties consistent with the observational constraints from GW170817, PSR J0740+6620, PSR J0030+0451, PSR J0437-4715, HESS J1731-347, and XTE J1814-338 \cite{arXiv:2409.15969,arXiv:2408.16583,arXiv:2402.14262}.
Including quark vector interactions and exchange interaction channels stabilizes nonstrange and strange quark matter and stiffens the equation of state (EoS). The resulting SQS models can reproduce masses, radii, and tidal deformabilities consistent with the observational inferences for PSR J0030+0451, PSR J0740+6620, and GW170817 \cite{arXiv:2203.04798}, demonstrating consistency with current observational constraints without uniquely identifying these objects as SQSs.
Models including repulsive interactions in strange quark matter have been shown to reproduce the observed properties of PSR J0740+6620 \cite{arXiv:2106.12956}.
The X-ray light curve of GRB 170714A has been interpreted as being consistent with a scenario involving the merger of two neutron stars, the temporary formation of a SQS, and its subsequent collapse into a black hole \cite{arXiv:1710.04355}.
The luminous supernova ASASSN-15lh has been proposed as a possible signature of the birth of a SQS \cite{arXiv:1508.07745}.
The long-duration gamma-ray burst GRB 240529A has been interpreted within a model involving the transition of a magnetar into a SQS, followed by SQS cooling and spin-down \cite{arXiv:2502.11511}.
A quark model including realistic vector interactions has also been proposed to describe compact objects such as the X-ray pulsar Her X-1 and the X-ray burster 4U 1820-30 within the SQS scenario \cite{arXiv:astro-ph/9810065}.

Despite extensive theoretical investigations, the existence of SQSs has not yet been confirmed observationally. Current measurements of compact objects, including mass-radius constraints from NICER and tidal deformability constraints from gravitational wave observations, provide important limitations on the EoS of dense matter but do not uniquely distinguish between hadronic neutron stars, hybrid stars, and self-bound SQSs. Nevertheless, several strange quark matter models can satisfy these observational constraints, motivating the study of SQSs as viable candidates for compact objects and as laboratories for investigating the properties of matter at supranuclear densities. Because phenomenological quark matter EoSs, including bag model descriptions, contain several free parameters, different parameter combinations may reproduce the currently available astrophysical constraints. Consequently, the current observational data are insufficient to establish that any observed compact star is a SQS or to uniquely determine the parameters of a preferred quark matter model. This motivates investigating additional observables that are sensitive not only to the EoS but also to the underlying theory of gravity. In particular, spontaneous scalarization introduces scalar charges and dipolar gravitational wave emission that may provide complementary information beyond the standard mass-radius and tidal-deformability measurements.

Different observations of compact objects have been employed to constrain the properties of strange quark matter and the corresponding EoSs.
Observational data from PSR J0740+6620 and GW170817 can constrain the parameters of MIT bag model EoSs used to describe strange quark matter \cite{arXiv:1909.00933}.
Measurements of the mass and radius of PSR J0030+0451 can be used to constrain the parameters of strange quark matter EoSs \cite{arXiv:2009.12571}.
Gravitational wave observations set constraints on the microscopic parameters of the bag model describing the quark matter \cite{arXiv:2409.11103}.
The self-binding of quark matter and the strength of repulsive quark interactions can be constrained using observational information on the maximum mass of compact stars \cite{arXiv:2503.11515}.
Under the assumption that old compact stars such as PSR J1801-0857D are SQSs, constraints on the scattering cross sections of light quarks and non-interacting scalar dark matter can be derived \cite{arXiv:1603.07518}.

Since SQSs provide a possible astrophysical realization of strange quark matter, the dependence of the
SQS properties on the various aspects of quark matter has been studied.
Nambu-Jona-Lasinio model \cite{arXiv:astro-ph/0101267,arXiv:nucl-th/0201003},
maximally symmetric phase of homogeneous superconducting quark matter
known as CFL phase \cite{arXiv:2007.00455},
density-dependent quark mass model \cite{arXiv:2408.15889}, and
color-spin-locked (CSL) quark matter \cite{arXiv:2409.15811} can be applied to explore
the structure of SQSs.
The scaling laws (relations between macroscopic observables and microscopic properties of stars) for SQSs can nevertheless be model-independent \cite{arXiv:2401.12519,arXiv:2503.11515}.
In contrast, the radial oscillation frequencies as well as the dynamical stability of SQSs are affected by the free parameter in the Vector MIT bag model \cite{arXiv:2412.05752}.
Bulk viscosity of strange quark matter and the damping time scales in SQSs depend on the quark model \cite{arXiv:astro-ph/0305320}.
Effective bag constant alters the radius, the mass, and the value of the moment of inertia in the SQSs \cite{arXiv:1601.06120}.
Mass-radius relation and the stability of SQS are affected by the fourth-order corrections parameter of the QCD perturbation
\cite{arXiv:1601.06120,arXiv:1903.03047,arXiv:2007.04121}.
Strong interaction effects in quark matter, i.e. perturbative QCD corrections and color superconductivity, can result
in a photon sphere and gravitational wave echoes \cite{arXiv:2107.09654}.
Isospin dependence of the quark mass affects the SQS mass-radius relation \cite{arXiv:1708.09599}.
The temperature dependence of the effective quark mass also affects the mass of SQSs and can modify their maximum mass during the heating and cooling evolution of proto-SQSs \cite{arXiv:1708.03428}.
A quark model incorporating a running coupling and a running strange-quark mass predicts stable strange-quark matter and SQS configurations with maximum masses around $2M_{\odot}$ \cite{arXiv:1512.08229}.
Depending on the strange-quark mass, SQSs in low-mass X-ray binaries may be accelerated to very high spin frequencies \cite{arXiv:astro-ph/0311128}.

The early evolution of SQSs, their strong magnetic fields and pressure anisotropy, as well as the effects of dark matter, electric charge, and electric fields, have been studied extensively.
In thermal emission from the surface of the hot SQSs, the emission of photons or electron-positron pairs can be dominated
depending on the surface temperature of the star \cite{arXiv:astro-ph/9712304,arXiv:astro-ph/0103361}.
In rotating SQSs, the star can absorb the crust matter thereby affecting the star cooling curve \cite{arXiv:astro-ph/0104116}.
Neutrino diffusion is the main process in the early cooling of SQSs \cite{arXiv:2110.00795}.
Strangelet crust alters the cooling and relaxation times of the SQSs \cite{arXiv:2201.06928}.
Theoretical models predict that strong magnetic fields can stiffen the EoS of strange quark matter, leading to more compact SQS configurations \cite{arXiv:1911.10512,arXiv:0812.0337}.
The transport coefficients and the instability region of SQSs are also influenced by the star magnetic field \cite{arXiv:0709.1224,arXiv:0812.0337,arXiv:0910.3633}.
The magnetic field of SQSs induces an anisotropic neutrino emission which results in the star's kick velocity \cite{arXiv:1801.06246}.
Anisotropy in the SQSs affects the mass-radius relation \cite{arXiv:2404.11723} as well as the radial oscillation and stability of these stars \cite{arXiv:1607.03984}.
In interacting SQSs with anisotropic pressure, the negative anisotropies enhance the star radial stability \cite{arXiv:2311.18770}.
Pressure anisotropy alters the frequencies and damping times of f-mode oscillations in anisotropic SQSs \cite{arXiv:2504.20589}.
Theoretical studies predict that, if SQSs exist, gravitational waves associated with glitch-induced f-mode oscillations may exhibit narrower distributions of f-mode frequencies than those of neutron stars, providing a potentially useful discriminant between different compact-star compositions \cite{arXiv:2403.09489}.
Theoretical studies have suggested that dark matter accretion may induce the conversion of compact stars into SQSs under suitable conditions \cite{arXiv:1007.1421}.
If SQSs exist, dark matter admixture may significantly influence their internal structure and global properties \cite{arXiv:1801.05031,arXiv:2505.20545,arXiv:2502.16317,arXiv:2409.09275}.
The interaction of dark matter with quark matter alters the f-mode oscillation in SQSs \cite{arXiv:2410.20923}.
The ultra-strong electric fields associated with color-superconducting strange quark matter can affect the mass and radius of SQSs \cite{arXiv:0907.5537}.
Theoretical studies have shown that electric charge can modify the structure and stability of SQS models \cite{arXiv:1509.07692,arXiv:2108.07454}.
In charged interacting SQS models, the stellar mass and radius increase with increasing electric charge \cite{arXiv:2006.00919,arXiv:2108.13972}.
Models of electrically charged SQSs predict that the f-mode gravitational wave frequencies depend sensitively on the stellar charge \cite{arXiv:2404.06412}.

Beyond these microphysical and astrophysical effects, the dependence of SQS properties on the underlying theory of gravity provides another important aspect of their theoretical modeling. Studies of SQS models in modified gravity have shown that their predicted masses and radii depend on the parameters in f(R) gravity \cite{arXiv:2206.03878,arXiv:2202.04467,arXiv:2112.09950,arXiv:2104.00590},
Horava gravity and Einstein-{\ae}ther theory \cite{arXiv:2006.07652}, Bumblebee gravity \cite{arXiv:2409.05801}, and Starobinsky gravity \cite{arXiv:1801.05031,arXiv:2006.06421}.
The non-minimal geometry-matter coupling theory has been shown to produce SQS models compatible with the observed properties of PSR J2215+5135, PSR J1614-2230, and the GW190814 event \cite{arXiv:2201.08726}.
Within massive gravity, SQS models have been proposed as possible candidates for compact objects in the mass-gap region $2.5-5 M_{\odot}$ \cite{arXiv:2402.14040}.
Within four-dimensional Einstein-Gauss-Bonnet gravity, studies of SQS models have shown that the allowed range of the bag constant \cite{arXiv:2110.03182}, as well as the predicted masses, radii \cite{arXiv:2108.07454}, and compactnesses \cite{arXiv:2006.00479}, depend on the gravitational coupling constant.
In the Starobinsky model, the total charge of the charged SQSs is also influenced by the gravitational background \cite{arXiv:2206.03878}.
Extra dimensions can alter the structure as well as the stability of SQSs \cite{arXiv:1907.07661}.

Among modified gravity theories, scalar-tensor gravity is particularly relevant to compact-star phenomenology because the additional scalar degree of freedom can modify stellar structure and introduce new gravitational observables. In a previous study, we employed the same scalar-tensor framework to investigate neutron stars containing nucleons, hyperons, and $\Delta$-resonances across different thermodynamic evolution regimes, and showed that the onset and properties of spontaneous scalarization depend on the composition and thermodynamic state of the stellar matter \cite{Rahimi2025}. The present work extends this line of investigation to self-bound SQSs, whose density profile, binding mechanism, and underlying EoS differ fundamentally from those of gravitationally bound neutron-star matter. Thus, although the scalar-tensor formalism is common to both studies, the stellar matter model and thermodynamic conditions considered here are fundamentally different. Scalar-tensor theories have also been constrained using X-ray pulse profiles of compact stars \cite{arXiv:2412.13867}, while scalar-field potentials and additional scalar interactions can modify the structure and stability of relativistic stars \cite{arXiv:2308.00203,arXiv:2107.13804,arXiv:1803.10510}.
In scalar-tensor gravity, spontaneous scalarization can be understood as a tachyonic instability of the scalar-free general-relativistic (GR) stellar configuration. When the scalar field is perturbed around the GR solution, its coupling to the matter distribution can effectively reduce the squared mass of the scalar perturbation. For sufficiently compact stellar configurations, this effective squared mass can become negative in part of the stellar interior, causing the scalar perturbation to grow rather than oscillate around the scalar-free state. The GR configuration then becomes unstable toward the development of a nonzero scalar field, and the system can settle into a new scalarized equilibrium configuration. The occurrence and strength of this instability depend on the stellar compactness and on the matter EoS, making the onset of spontaneous scalarization sensitive to the properties of the stellar interior.
Couplings of the scalar field to the U(1) gauge field \cite{arXiv:2105.14661} and to the Gauss-Bonnet invariant \cite{arXiv:1711.02080} can also lead to spontaneous scalarization in charged compact stars.
With a massive scalar field, the dynamical transition to spontaneous scalarization may occur in compact stars \cite{arXiv:2407.08124}.
Self-interacting scalar fields in Einstein-Gauss-Bonnet gravity, with coupling between the scalar field and the Gauss-Bonnet term, can evolve during stellar collapse \cite{arXiv:1712.05149}.
The coupling of the scalar field to the Gauss-Bonnet invariant can induce scalar hair in compact stars \cite{arXiv:1910.13718}.
Moreover, couplings between the scalar field and the Gauss-Bonnet invariant, as well as between the scalar field and curvature invariants such as the Ricci scalar, can affect the domain of stellar existence and the scalar charge of the star \cite{arXiv:2111.03644}.
Applying data from the neutron-star--black-hole binary GW200115, the neutron-star scalar charge has been constrained \cite{arXiv:2311.09281}.
Universal relations between the scalar charge of scalarized stars and their binding energy \cite{arXiv:2105.01614}, as well as between the moment of inertia and quadrupole moment in rotating scalarized stars \cite{arXiv:1408.1641}, have been found to be independent of the stellar EoS.
In some compact objects, scalar clouds can extend beyond the stellar radius and contribute to the Arnowitt-Deser-Misner (ADM) mass of the star \cite{arXiv:2312.15839}.
Scalarized stars are also affected by the mass of the gravitational scalar and can have masses and radii different from those of their non-scalarized counterparts \cite{arXiv:2310.05200}.

As discussed above, current astrophysical observations constrain the macroscopic properties of compact stars but do not uniquely determine their internal composition. Consequently, several hadronic, hybrid-star, and SQS EoSs remain compatible with existing measurements. The dependence of scalarization on the composition and thermodynamic state of neutron-star matter demonstrated in our previous study \cite{Rahimi2025} naturally raises the question of whether analogous microphysics-dependent scalarization occurs in self-bound SQSs, whose EoS and density profile differ qualitatively from those of gravitationally bound neutron-star matter. Since spontaneous scalarization is highly sensitive to stellar compactness and the EoS, EoS-dependent scalar charges and the associated dipolar gravitational wave emission may provide complementary observables for distinguishing SQSs from conventional neutron stars while simultaneously constraining scalar-tensor gravity. In the present work, we therefore investigate SQSs in scalar-tensor gravity, focusing on how different strange quark matter EoSs affect the onset of spontaneous scalarization and the resulting scalar charge. The rest of our paper is structured as follows. In Section II, we explain the EoSs of strange quark matter based on the bag models employed in this work. Section III describes the scalar-tensor gravity and the modified structural equations for relativistic stars. In Section IV, we present our results for the scalarized SQSs. Section V gives the Summary and Conclusions.

\section{Strange quark matter equations of state}\label{s2}

In this study, we apply three EoSs for the strange quark matter based on the bag models. We assume the superfluid strange quark matter contains the massless particles (u, d quarks, and electrons), as well as s quarks with finite mass $m_s$ \cite{Haensel1986}. These three cases are normal quark matter, quark matter in the CFL phase, and quark matter in the CFLm model. In the bag model, the grand canonical potential per unit volume for the normal quark matter is as follows,
\begin{eqnarray}
          \Omega_{Normal}=\sum_{i=u,d,s,e} \Omega_i^0+\frac{3(1-a_4)}{4\pi^2}\mu^4+B_{eff},
 \end{eqnarray}
in which $\Omega_i^0$ presents the grand canonical potential for particle type $i$ as the ideal Fermi gas \cite{Farhi}. Furthermore,
$\mu=(\mu_u+\mu_d+\mu_s)/3$ is the average quark chemical potential, the effective bag constant, $B_{eff}$, denotes the contributions from the quantum chromodynamics (QCD) vacuum,
and $a_4$ gives the perturbative QCD contribution from one-gluon exchange for gluon interaction.
The smaller value of $a_4$ means a larger correction, since the strength of the correction is ($1-a_4$).

In addition, the number density for each part, $n_i$, relates to the chemical potential, $\mu_i(i=u,d,s,e)$, as follows,
\begin{eqnarray}
           n_i=-\frac{\partial\Omega}{\partial\mu_i}.
 \end{eqnarray}
Considering the quark matter at the equilibrium state, the weak interactions result in the conditions,
\begin{eqnarray}
          \mu_d=\mu_u+\mu_e,
 \end{eqnarray}
\begin{eqnarray}
          \mu_d=\mu_s.
 \end{eqnarray}
Moreover, the charge neutrality leads to the condition,
\begin{eqnarray}
         \frac{2}{3}n_u= \frac{1}{3}[n_d+n_s]+n_e.
 \end{eqnarray}
In the case of normal quark matter, the pressure of quark matter is given by,
\begin{eqnarray}
        P_{Normal}=- \Omega_{Normal},
 \end{eqnarray}
with the following quark matter energy density,
\begin{eqnarray}
        \varepsilon_{Normal}= \Omega_{Normal}+\sum_{i=u,d,s,e} \mu_i n_i.
 \end{eqnarray}
In the second model for the superfluid quark matter known as Color-Flavor Locked (CFL) model, an additional term describing the
pairing energy is added to the Normal grand canonical potential \cite{Li},
\begin{eqnarray}
          \Omega_{CFL}= \Omega_{Normal}+\frac{3m_s^4-48\Delta^2\mu^2}{16\pi^2},
 \end{eqnarray}
in which $\Delta$ denotes the uncertain CFL pairing gap. In the above equation, the second term characterizes the condensation energy particular to the CFL phase. This term adds an additional repulsive component that stiffens the EoS. The third model describing the superfluid quark matter is CFLm model in which
the strange quark mass $m_s$ is considered as a free parameter.
\begin{figure}[h]
	\subfigure{}{\includegraphics[scale=0.7]{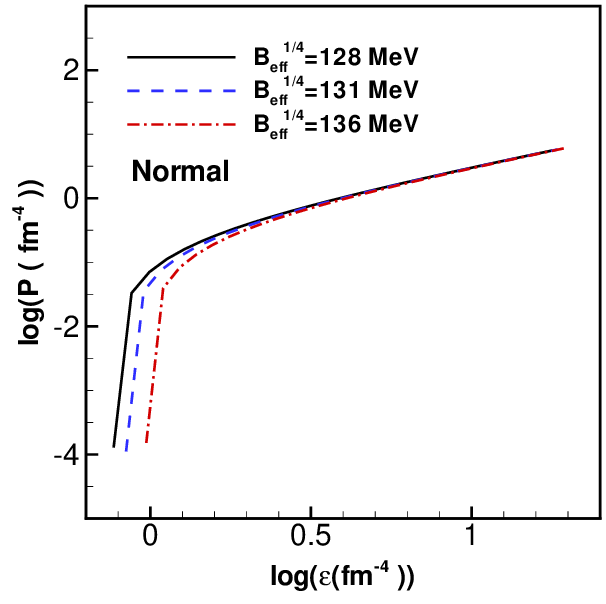}
		}	
\subfigure{}{\includegraphics[scale=0.7]{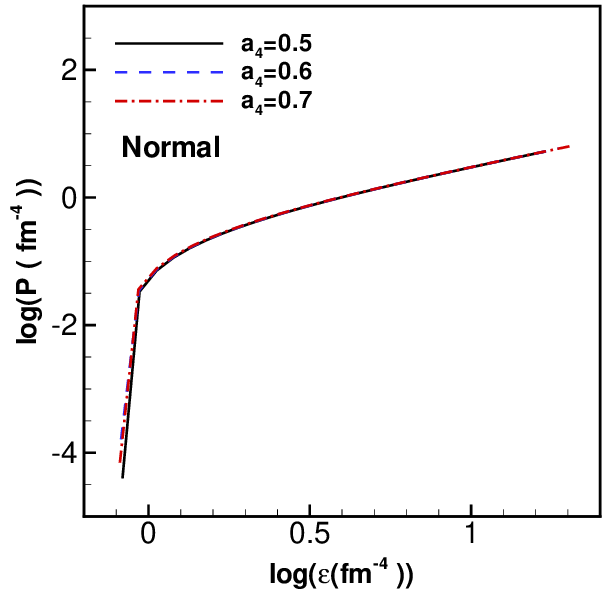}
		}
	\caption{Equation of state (logarithmic scale) of strange quark matter in the Normal model for a fixed strange-quark mass $m_s=100$ MeV. Left: $a_4=0.58$ and different values of $B_{\rm eff}$; right: $B_{\rm eff}^{1/4}=130.3$ MeV and different values of $a_4$. These panels show clearly the differences among the EoS parameterizations in the low-pressure regime.}
	\label{pNormal}
\end{figure}

\begin{figure}[h]
	\subfigure{}{\includegraphics[scale=0.7]{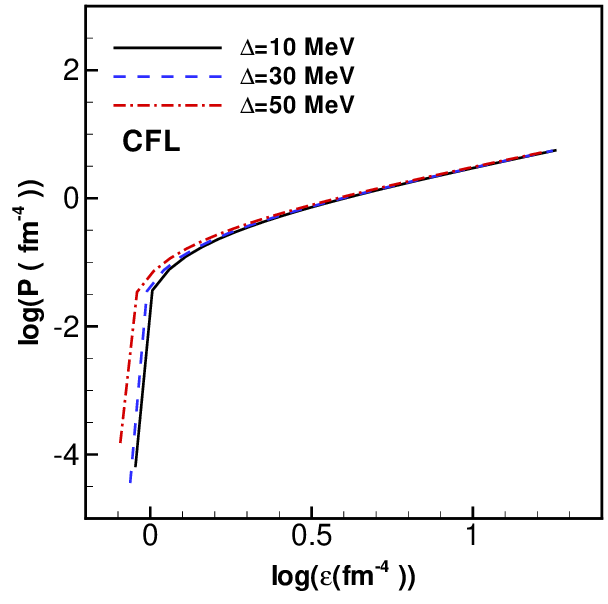}
		}	
	\caption{Equation of state (logarithmic scale) of strange quark matter in the CFL model for $m_s=100$ MeV, $a_4=0.53$, and $B_{\rm eff}^{1/4}=133.1$ MeV, considering different values of the CFL pairing gap $\Delta$. It clearly displays the differences among the EoS parameterizations in the low-pressure regime and the corresponding finite energy density at the stellar surface.}
	\label{pCFL}
\end{figure}

\begin{figure}[h]
	\subfigure{}{\includegraphics[scale=0.7]{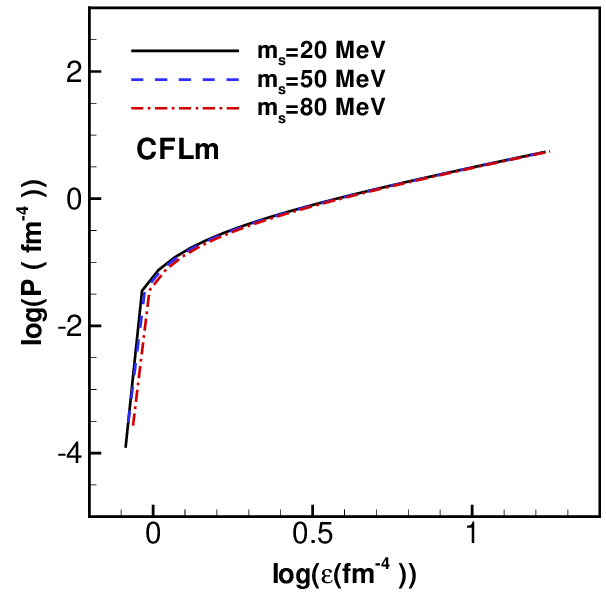}
		}	
	\caption{Equation of state (logarithmic scale) of strange quark matter in the CFLm model for $a_4=0.53$, $\Delta=33.9$ MeV, and $B_{\rm eff}^{1/4}=134.4$ MeV, considering different values of the strange-quark mass $m_s$. It displays the differences among the EoS parameterizations in the low-pressure regime and the corresponding finite energy density at the stellar surface.}
	\label{pCFLm}
\end{figure}

Figures \ref{pNormal}-\ref{pCFLm} present the EoS of strange quark matter in three models considered in this work. Figure \ref{pNormal} verifies that for Normal quark matter, the higher values of the effective bag constant, $B_{eff}$, soften the EoS. Besides, increasing $B_{\rm eff}$ shifts the low-pressure point toward higher energy density. In our numerical results, the energy density at the lowest-pressure point increases from approximately $0.769$ to $0.972~{\rm fm}^{-4}$ as $B_{\rm eff}^{1/4}$ increases from $128$ to $136$ MeV. By contrast, increasing $a_4$ produces a comparatively smaller shift in the low-pressure energy density, which decreases from approximately $0.830$ to $0.815~{\rm fm}^{-4}$ for $a_4=0.5$--$0.7$. However, the strange quark matter EoS becomes stiffer by increasing the perturbative QCD contribution in gluon interaction, $a_4$. The rate at which pressure increases with $a_4$ grows with energy density. Moreover, $a_4$ slightly reduces the surface density.
Therefore, $B_{eff}$ and $a_4$ have opposite effects on SQSs. However, the influence of the effective bag constant on the EoS is more significant than the $a_4$ ones.
We can realize from Figure \ref{pCFL} that the EoS of strange quark matter in CFL phase gets stiffer by growing the uncertain CFL pairing gap, $\Delta$.
In fact, the higher values of $\Delta$ affect the condensation energy leading to the stiffer EoS. The effects of the uncertain CFL pairing gap on the quark matter are also more notable at higher densities. Besides, increasing $\Delta$ decreases the energy density corresponding to the low-pressure surface. Specifically, it decreases from approximately $0.899$ to $0.807~{\rm fm}^{-4}$ as $\Delta$ increases from $10$ to $50$ MeV. Thus, the increase of the CFL pairing gap not only stiffens the EoS but also shifts the low-pressure endpoint toward a lower energy density.
Figure \ref{pCFLm} confirms that considering CFLm model for strange quark matter, an increase in the strange quark mass, $m_s$, results in a softening of the EoS, with the most pronounced effects occurring at higher densities. This stems from the fact that heavier strange quarks produce a smaller Fermi pressure.
In addition, increasing the strange-quark mass $m_s$ shifts the low-pressure endpoint toward higher energy density. The corresponding energy density increases from approximately $0.820$ to $0.865~{\rm fm}^{-4}$ as $m_s$ increases from $20$ to $80$ MeV. Therefore, $m_s$ and $B_{\rm eff}$ both increase the energy density associated with the low-pressure surface, whereas $\Delta$ and $a_4$ produce the opposite trend over the parameter ranges considered here. As can be seen from Figures \ref{pNormal}-\ref{pCFLm},
both $B_{eff}$ and $m_s$ are the parameters which soften the SQS EoS, with a more considerable impact of $B_{eff}$ compared to $m_s$.
This is while the other two parameters $a_4$ and $\Delta$ lead to the stiffening of the EoS, with $\Delta$ exerting a greater influence than $a_4$. These results demonstrate that the different EoS models and their parameters lead to distinct finite energy densities in the low-pressure limit. The described EoS characteristics are the key determinants of how the stellar matter EoS drives spontaneous scalarization in SQSs. In the following of the paper, we are interested in how the properties of strange quark matter affect the star scalarization.

\section{Compact stars in scalar-tensor theories of gravity}\label{s}

In order to explore the spontaneous scalarization in SQSs, we employ the formalism describing the compact objects in scalar-tensor gravity
introduced in Ref. \cite{arXiv:1604.04175}. Considering the Einstein frame, the action in this gravity is as follows,
	\begin{equation} \label{action}
		S[g_{\mu \nu},\Phi,\Psi_m] = \frac{1}{16 \pi} \int d^4x \sqrt{-g}(R-2 \nabla_\mu\Phi \nabla^\mu \Phi)+S_m[\Psi_m,a(\Phi)^2 g_{\mu \nu}].
	\end{equation}
Here, $g_{\mu \nu}$ is the Einstein metric with $g=det(g_{\mu \nu})$, $R$ presents the Ricci scalar, $\Phi$ denotes the scalar field, and $\Psi_m$ is the matter field. Besides, $a(\Phi)$ describes the coupling function and $\tilde{g} _{\mu \nu}=a(\Phi)^2 g_{\mu \nu}$ is the Jordan metric. The variation of Eq. (\ref{action}) results in the field equations,
\begin{equation} \label{E1}
	G_{\mu \nu}-2 \nabla_\mu\Phi \nabla_\nu \Phi+g_{\mu \nu}\nabla_\rho\Phi \nabla^\rho \Phi=8\pi a^2 \tilde{T}_{\mu \nu},
\end{equation}
\begin{equation} \label{E2}
	\nabla^\mu \nabla_\mu\Phi=-4\pi a^4 \alpha \tilde{T},
\end{equation}
in which $\tilde{T}_{\mu \nu}$ is the stress-energy-momentum tensor of the matter fields, $\tilde{T}=\tilde{g} _{\mu \nu}\tilde{T}^{\mu \nu}$, and $\alpha(\Phi)=\frac{d ln a(\Phi)}{d \Phi}$. Here, we investigate the Einstein frame metric considering the fluid in the Jordan frame. In this work, following the assumptions in Ref. \cite{arXiv:1604.04175}, we consider the form $a(\Phi) = e^{\frac{1}{2}\beta (\Phi-\Phi_0) ^2}$ in which $\beta$ is the coupling constant and $\Phi_0=0$.

For the coupling function adopted here, $\alpha(\Phi)=\beta(\Phi-\Phi_0)$. Consequently, the scalar-free configuration,
$\Phi=0$, is always an exact solution of the field equations. Its
stability, however, depends on the coupling parameter $\beta$ and on
the properties of the stellar matter. This can be seen by considering
a small perturbation about the scalar-free configuration,
$\Phi=\delta\Phi$. To linear order, one has
$a(\Phi)\simeq 1$ and
$\alpha(\Phi)\simeq\beta\delta\Phi$. The scalar field equation therefore
reduces to
\begin{equation}
\nabla^\mu\nabla_\mu \delta\Phi
=
-4\pi\beta\tilde{T}\,\delta\Phi,
\label{eq:linear_scalar}
\end{equation}
or, equivalently,
\begin{equation}
\left(
\nabla^\mu\nabla_\mu
-
m_{\rm eff}^2
\right)\delta\Phi
=
0,
\qquad
m_{\rm eff}^2
=
-4\pi\beta\tilde{T},
\label{eq:effective_mass}
\end{equation}
where
\begin{equation}
\tilde{T}
=
-\tilde{\epsilon}+3\tilde{p}
\label{eq:trace_definition}
\end{equation}
is the trace of the matter energy-momentum tensor in the Jordan frame. A negative effective mass squared, $m_{\rm eff}^2<0$, over a
sufficiently extended region of the stellar interior signals a
tachyonic instability of the scalar-free solution. For the stellar
matter considered in this work, the trace is generally negative over
the relevant density range, $\tilde{T}<0$. Hence, the condition
$m_{\rm eff}^2<0$ is naturally associated with negative values of
$\beta$. Moreover, increasing the magnitude of a negative $\beta$
makes $m_{\rm eff}^2$ more negative and consequently strengthens the
tachyonic instability. This provides the physical motivation for
focusing on the negative-$\beta$ sector in the present analysis.

For positive values of $\beta$, since $\tilde{T}<0$ for the stellar matter considered here, the effective mass squared remains positive, $m_{\rm eff}^2>0$, and therefore the tachyonic mechanism responsible for spontaneous scalarization is absent. Consequently, the scalar-free solution remains stable and no spontaneous scalarization is expected in this regime. In contrast, as $\beta$ becomes increasingly negative, $m_{\rm eff}^2$ becomes progressively more negative, strengthening the tachyonic instability and causing the onset of scalarization to occur at lower central densities. Thus, increasingly large $|\beta|$ on the negative-$\beta$ side corresponds qualitatively to progressively stronger scalarization. We do not extend the numerical analysis to arbitrarily large negative values of $\beta$, since the values $\beta=-4.5$, $-5$, and $-6$ already cover the near-threshold to strongly scalarized regimes relevant to the SQS configurations investigated in this work.

The GR limit is obtained for $\beta=0$. In this
case, $\alpha(\Phi)=0$ for all $\Phi$, and the scalar field is
completely decoupled from the matter sector. Imposing the asymptotic
condition $\Phi(r\rightarrow\infty)=0$, the regular equilibrium
solution is then $\Phi=0$, and the stellar structure equations reduce
exactly to those of general relativity. For weak negative coupling,
the scalar-free configuration remains stable until the stellar
compactness and matter distribution become sufficiently large for the
tachyonic instability to develop. Thus, spontaneous scalarization
occurs only when $\beta$ becomes sufficiently negative, with the
critical value depending on the EoS and on the
corresponding stellar sequence.

For the class of scalar-tensor theories considered here, spontaneous
scalarization is known to occur for sufficiently negative coupling
parameters, with a representative onset around
$\beta\simeq -4.35$ for neutron-star configurations. This value should
not, however, be interpreted as a universal critical coupling, since
the precise threshold depends on the EoS, stellar
compactness, and the associated equilibrium sequence. Accordingly,
we do not regard $\beta=-4.5$ as a universal critical value. Instead,
we adopt it as a representative coupling close to the onset of
scalarization, while $\beta=-5$ and $\beta=-6$ probe progressively
stronger scalarization. This choice allows us to follow the evolution from
the near-threshold regime to increasingly strong scalarization while keeping the number of representative coupling values manageable.

Linear perturbations of the scalar field around the scalar-free configuration couple to the trace of the matter stress-energy tensor and can acquire an effective negative squared mass in sufficiently compact stellar interiors. This is the tachyonic-type instability responsible for the onset of spontaneous scalarization. When the scalar-free solution becomes unstable, the system can develop a nonzero scalar field and approach a scalarized equilibrium configuration. In the present work, we identify scalarized solutions by solving the full nonlinear generalized TOV equations and imposing the asymptotic condition $\Phi(r\rightarrow\infty)=0$, rather than by performing a separate time-dependent perturbation analysis. The coupling parameter $\beta$ controls the strength of the scalar-matter coupling and, together with the stellar compactness and the EoS, determines the susceptibility of the scalar-free configuration to spontaneous scalarization.

For the spherically symmetric perfect fluid in SQSs, the stress-energy-momentum tensor is,
\begin{equation}
	\tilde{T}^{\mu \nu}=\tilde{\epsilon}\tilde{u}^{\mu}\tilde{u}^{\nu}+\tilde{p}(\tilde{g}^{\mu \nu}+\tilde{u}^{\mu}\tilde{u}^{\nu}),
\end{equation}
with the energy density $\tilde{ \epsilon}$ and pressure $\tilde{p}$ of the fluid with 4-velocity $\tilde{\textbf{u}}$ and EoS of $\tilde{p}(\tilde{ \epsilon})$.
The line element of the space-time for the spherically symmetric static star is,
\begin{equation}
	ds^2 = - N(r)^2 dt^2 + A(r)^2 dr^2 + r^2 (d\theta^2 + \sin^2\theta d\varphi^2),
\end{equation}
with the metric functions $N(r)$ and $A(r) = [1-2 m(r)/r ]^{-1/2}$ in which $m(r)$ is the mass profile. Eqs. (\ref{E1}) and (\ref{E2}) in the static limit lead
to the generalized Tolman-Oppenheimer-Volkoff (TOV) equations \cite{arXiv:1604.04175},
	\begin{align}
		&\frac{d m}{dr} = 4\pi r^2 a^4 \tilde{\epsilon} + \frac{r}{2} (r-2m) \Big(\frac{d\Phi}{dr}\Big)^2 \label{eq:dm},\\
		&\frac{d \ln N}{dr} = \frac{4\pi r^2 a^4 \tilde{p}}{r - 2m} +\frac{r}{2} \Big(\frac{d\Phi}{dr}\Big)^2 + \frac{m}{r(r-2m)} \label{eq:dn}, \\
		&\frac{d^2\Phi}{dr^2} = \frac{4\pi r a^4}{r-2m} \! \left[ \alpha (\tilde{\epsilon} - 3\tilde{p}) + r (\tilde{\epsilon} - \tilde{p}) \frac{d\Phi}{dr} \right ]\! -\frac{2(r-m)}{r(r-2m)} \frac{d\Phi}{dr} \label{eq:dphi}, \\
		&\frac{d\tilde{p}}{dr} = -(\tilde{\epsilon} + \tilde{p}) \left[  \frac{4\pi r^2 a^4 \tilde{p}}{r-2m} \! + \! \frac{r}{2} \Big(\frac{d\Phi}{dr}\Big)^2 \!\! + \! \frac{m}{r(r-2m)} \! + \! \alpha \frac{d\Phi}{dr} \right]. \label{eq:dp}
			\end{align}
To solve the above equations, we should apply the boundary conditions as follows,
\begin{align}
		&m(0)= 0, \quad \lim_{r\to\infty}N(r) = 1,\quad \Phi(0)=\Phi_c, \quad \lim_{r\to\infty}\Phi(r) = 0, \nonumber \\
		&\dfrac{d\Phi(r)}{dr}(0) = 0, \qquad \tilde{p}(0) = p_c, \qquad \tilde{p}(R_s) = 0,    \label{eq:bc}
	\end{align}
in which $R_s$ presents the star radius and $c$ denotes the center of the star. To solve the generalized TOV equations, we should match the solutions in the stellar interior with the exterior analytical solutions at the surface of the star. Considering a guess for $\Phi(0)=\Phi_c$ at the star center and an iteration on $\Phi_c$ to satisfy the below condition at the surface of the star,
\begin{equation} \label{eq:constraint on central of scalar field}
		\Phi_s  + \frac{2 \psi_s}{\sqrt{\dot{\nu}_s^2+4\psi_s^2}} \textrm{arctanh} \left[ \frac{\sqrt{\dot{\nu}_s^2 +4\psi_s^2}}{\dot{\nu}_s +2/R_s} \right] = 0,
	\end{equation}
lead to the solutions of the equations \cite{26,arXiv:1604.04175}. Here, $\psi_s = (d\Phi/dr)_s$ and $\dot{\nu}_s = 2(d\ln N/dr)|_s = R_s \psi_s^2 + 2 m_s/[R_s(R_s-2m_s)]$. The ADM mass, i.e. the total gravitational star mass measured from spatial infinity, is obtained from the general exterior solution,
\begin{align}
		M_{ADM} &= \frac{R_s^2 \dot{\nu}_s}{2} \left( 1-\frac{2m_s}{R_s} \right)^\frac{1}{2}
		\exp \left[ \frac{-\dot{\nu}_s}{\sqrt{\dot{\nu}_s^2+4\psi_s^2}} \textrm{arctanh} \left( \frac{\sqrt{\dot{\nu}_s^2+4\psi_s^2}}{\dot{\nu}_s +2/R_s} \right) \right].
	\end{align}
Furthermore, the scalar charge, $\omega$, is related to ADM mass as follows,
	\begin{align} \label{omM}
				\omega & = - 2 M_{ADM} \psi_s/\dot{\nu}_s.
	\end{align}

The scalar charge, $\omega$, characterizes the asymptotic strength of
the scalar field and is determined by the complete scalar field profile
inside and outside the star. In particular, $\omega$ is not identical to
the central scalar field $\Phi_c$ and is not expected to be universally
proportional to $\Phi_c$. Rather, for a given EoS and
coupling constant $\beta$, the central value $\Phi_c$ fixes the regular
interior solution of the scalar field equation, and this solution
determines the scalar field derivative at the stellar surface,
$\psi_s=(d\Phi/dr)_s$. Through the matching to the exterior solution,
$\psi_s$ and the corresponding metric quantities determine the scalar
charge according to Eq.~(\ref{omM}). Thus, there is a one-to-one
correspondence between the central scalar field and the asymptotic
scalar charge along a given equilibrium sequence, although the precise
mapping $\omega(\Phi_c)$ depends on the EoS and on the
coupling constant.

In the weakly scalarized regime, $\omega$ increases approximately with
$\Phi_c$, while in the strongly scalarized regime the relation can become
nonlinear and may exhibit saturation. Therefore, a larger central scalar
field generally corresponds to a stronger exterior scalar charge, but
the magnitude of $\omega$ also depends on the stellar compactness and
the radial distribution of the scalar field. For the asymptotically flat solutions considered here, the GR branch
corresponds to $\Phi(r)=0$ everywhere and hence $\omega=0$, whereas a
scalarized SQS has a nonzero scalar charge.

\section{Results and Discussions}\label{s}

\subsection{Strange Quark Star Mass}\label{s}

\begin{figure}[h]
	\subfigure{}{\includegraphics[scale=0.36]{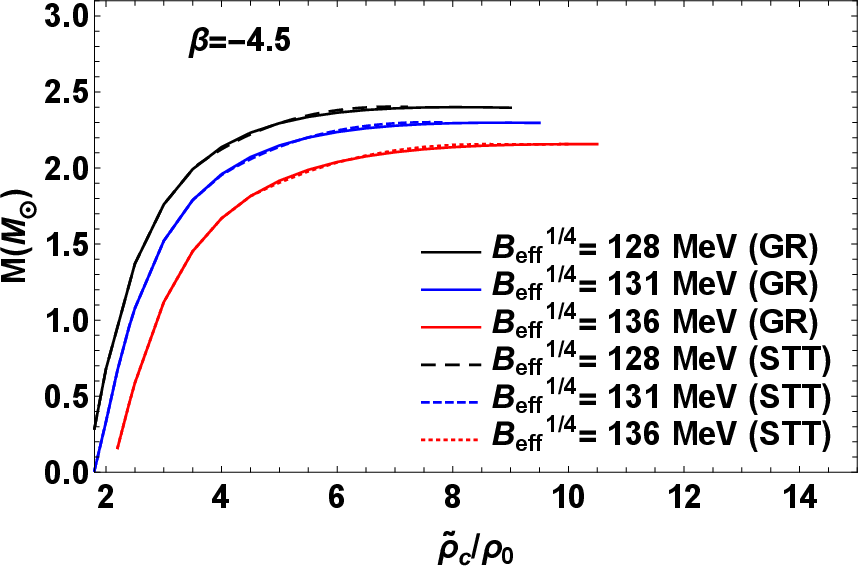}
		}	
\subfigure{}{\includegraphics[scale=0.36]{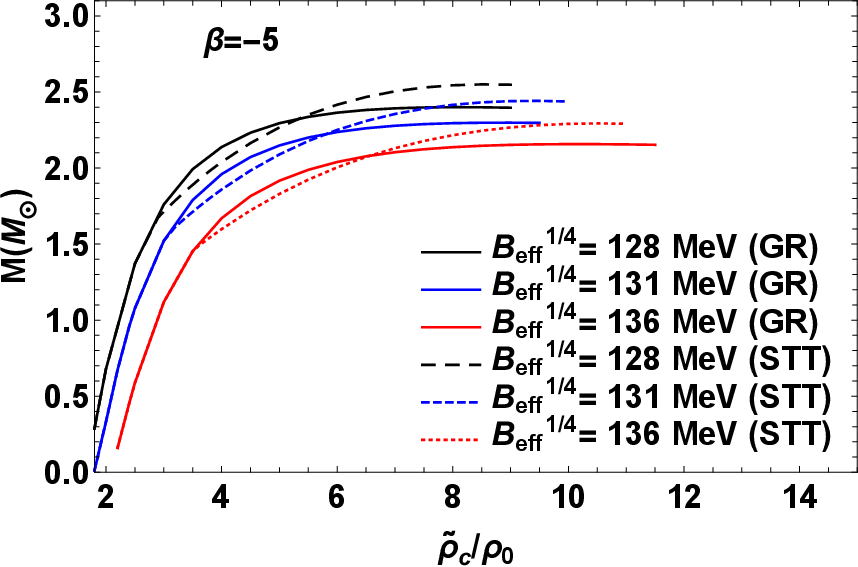}
		}
\subfigure{}{\includegraphics[scale=0.36]{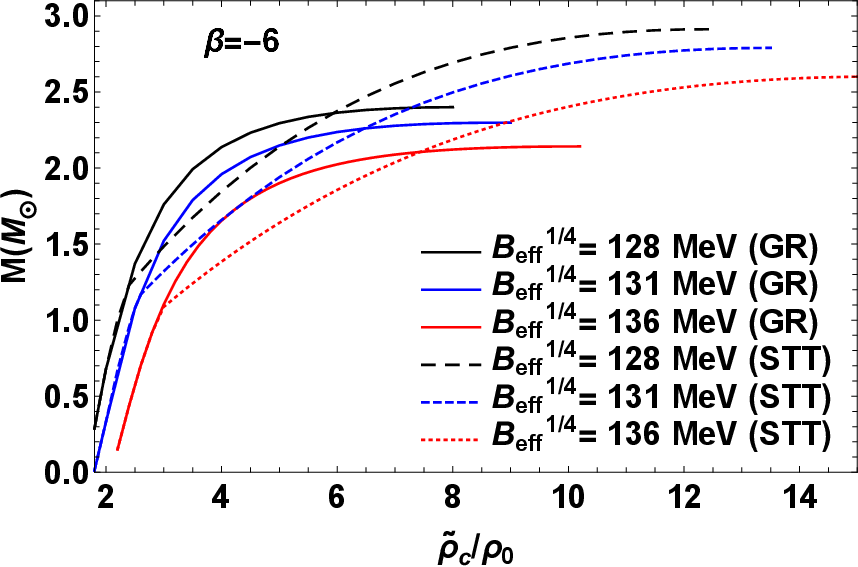}
		}\\
	\subfigure{}{\includegraphics[scale=0.36]{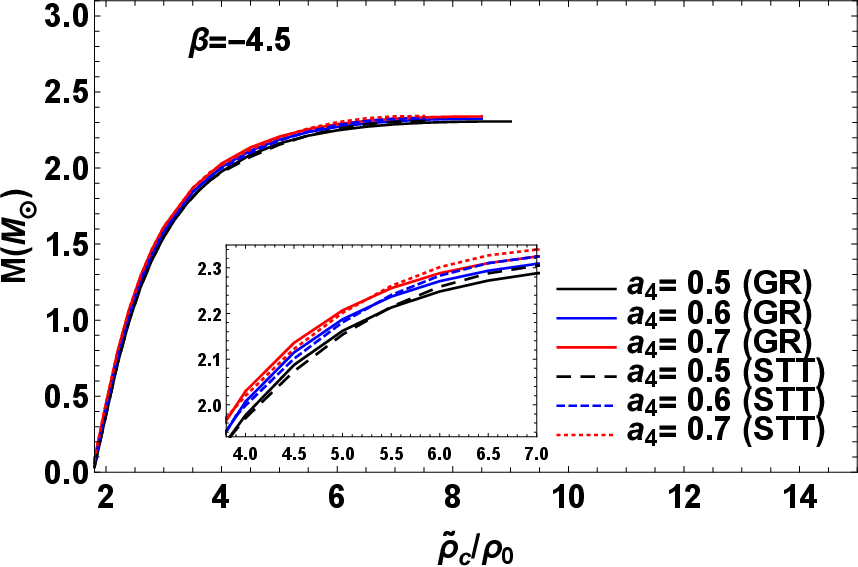}
		}	
\subfigure{}{\includegraphics[scale=0.36]{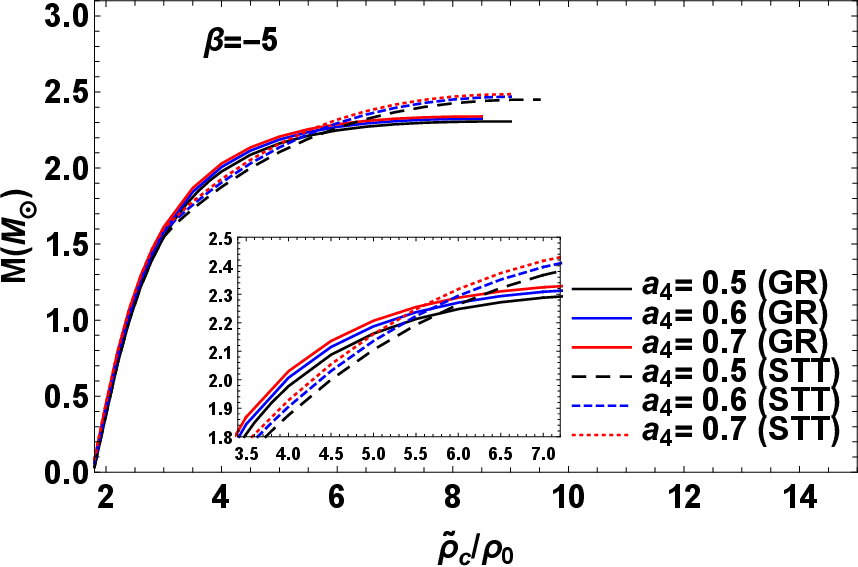}
		}
\subfigure{}{\includegraphics[scale=0.36]{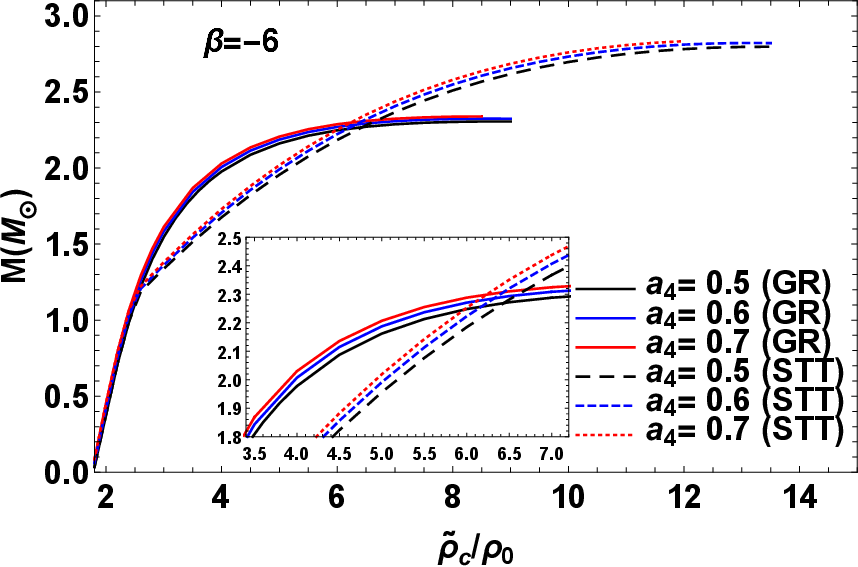}
		}
	\caption{Strange quark star mass versus the central density in Normal model for a fixed strange quark mass $m_s=100MeV$, Top: considering the value of $a_4=0.58$ and different values of $B_{eff}$, and Bottom: considering the value of $B_{eff}^{1/4}=130.3MeV$ and different values of $a_4$, for three values of the coupling constant, $\beta$, in the scalar-tensor theory (STT) and general relativity (GR). The value $\rho_0=1.66\times10^{14}g/cm^3$ is applied to present the dimensionless density.}
	\label{MroNormal}
\end{figure}

\begin{figure}[h]
	\subfigure{}{\includegraphics[scale=0.36]{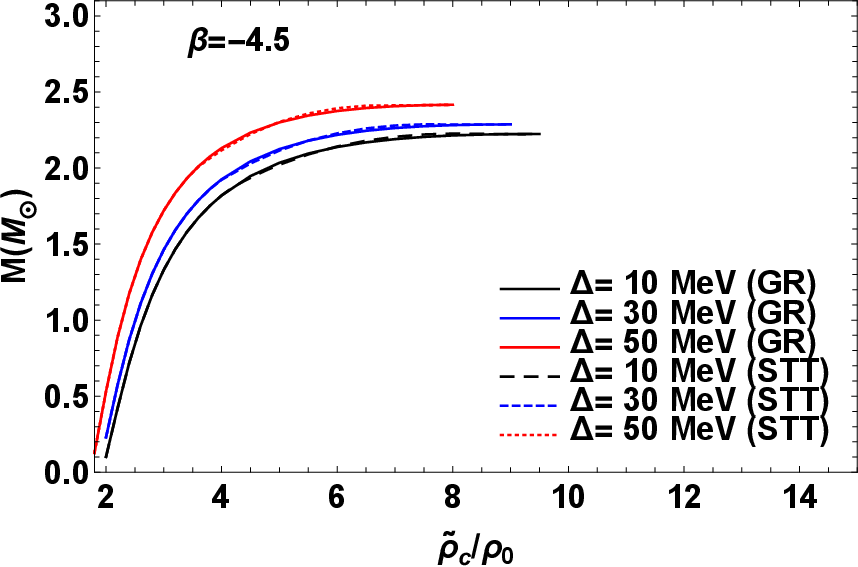}
		}	
	\subfigure{}{\includegraphics[scale=0.36]{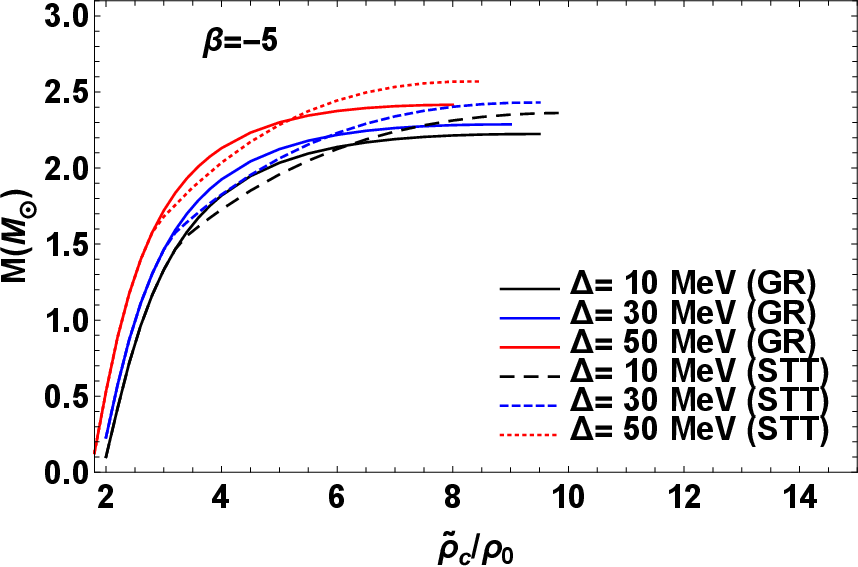}
		}
	\subfigure{}{\includegraphics[scale=0.36]{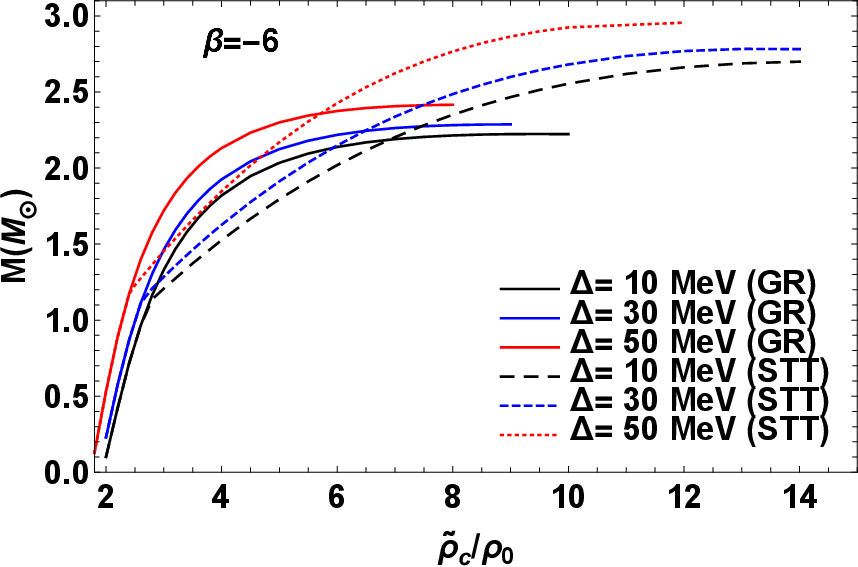}
		}
	\caption{Strange quark star mass versus the central density considering CFL model with the values of $m_s=100MeV$, $a_4=0.53$, $B_{eff}^{1/4}=133.1MeV$, and different values of $\Delta$ for three values of the coupling constant, $\beta$, in STT and GR.}
	\label{MroCFL}
\end{figure}

\begin{figure}[h]
	\subfigure{}{\includegraphics[scale=0.36]{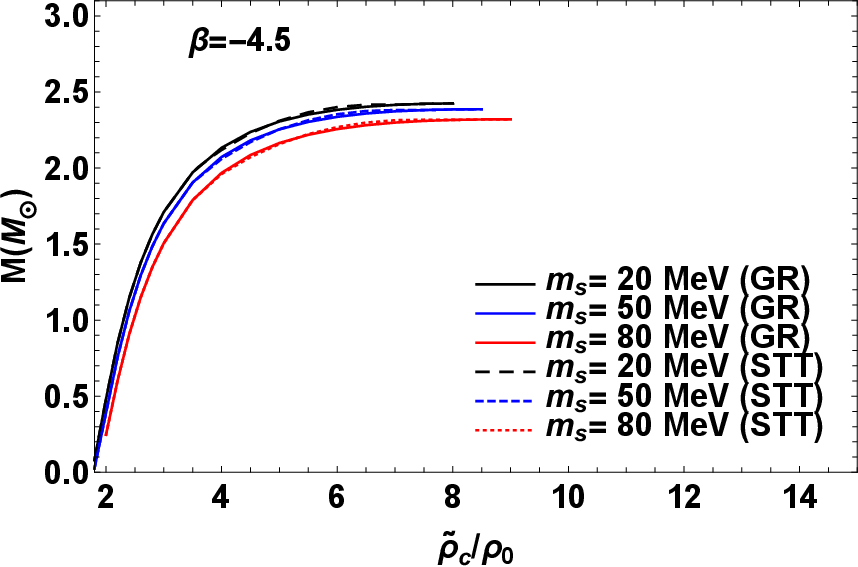}
		}	
	\subfigure{}{\includegraphics[scale=0.36]{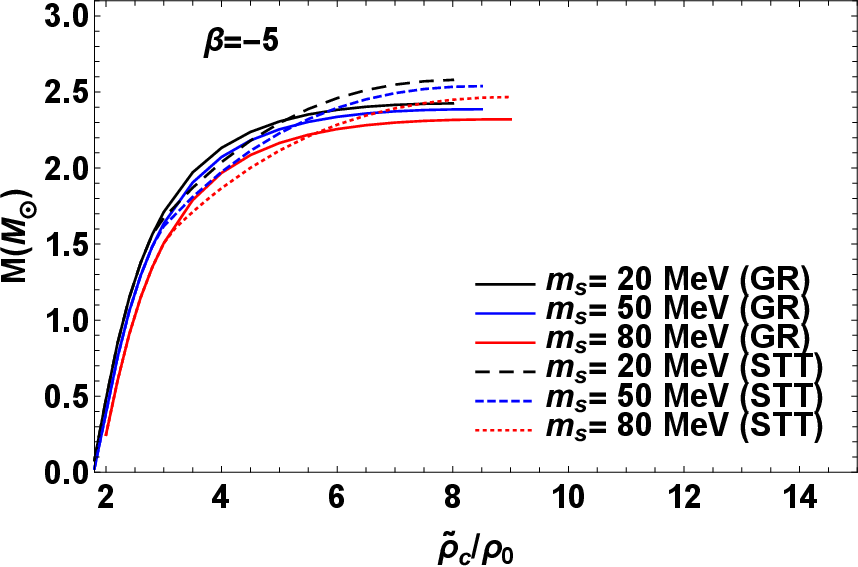}
		}
	\subfigure{}{\includegraphics[scale=0.36]{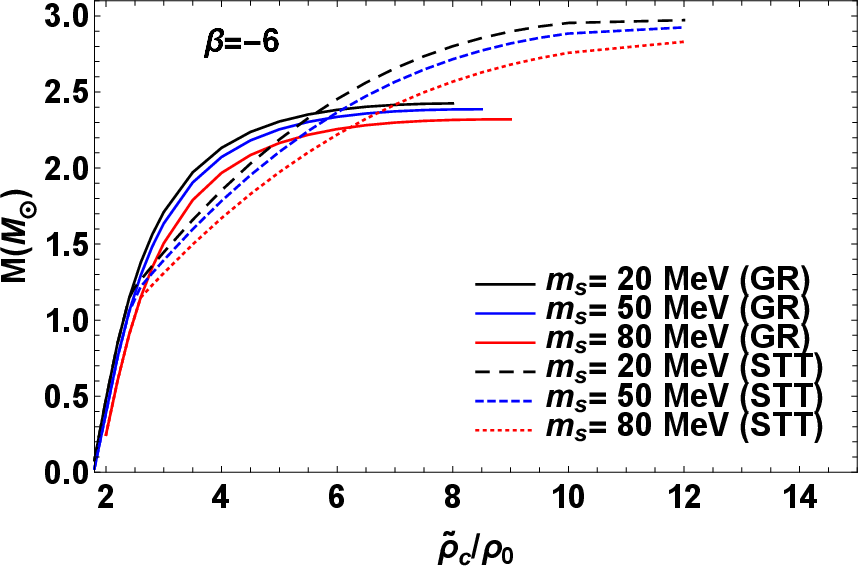}
		}
	\caption{Strange quark star mass versus the central density in CFLm model with the values of $a_4=0.53$, $\Delta=33.9MeV$, $B_{eff}^{1/4}=134.4MeV$, and different values of $m_s$ for three values of the coupling constant, $\beta$, in STT and GR.}
	\label{MroCFLm}
\end{figure}

Here, we report our results for the ADM mass of SQSs. Figures \ref{MroNormal}-\ref{MroCFLm} show the SQS mass versus the central density in different strange quark matter models. These Figures verify that in scalar-tensor gravity, spontaneous scalarization can alter the sequence of hydrostatic equilibrium solutions by
developing a non-trivial scalar field that modifies the pressure support. This causes the mass of the scalarized star to deviate from the GR value. The degree to which the mass of a scalarized SQS is enhanced or suppressed relative to the GR prediction depends on the central density. Besides, for the SQSs with the moderate coupling $\beta=-4.5$, spontaneous scalarization manifests as a slight effect in the mass plot. Considering the couplings $\beta=-5$ and $\beta=-6$,
the tachyonic instability sets in at a lower central density than for $\beta=-4.5$. Moreover, for the lower values of
the coupling $\beta$, the scalarized branch separates from the GR one over a much wider range and the maximum mass is significantly enhanced. For $\beta=-6$, the scalarized branch extends over the entire astrophysically relevant density range, making these SQSs strongly scalarized objects.

Figure \ref{MroNormal} indicates that for Normal SQSs, increasing $B_{eff}$ which softens the EoS yields a lower peak mass and delays the onset of scalarization to higher central densities. This behavior arises because a softer EoS reduces compactness and gravitational potential, weakens the tachyonic instability, and a higher central density is needed to trigger scalarization. The fact that even the softest EoS ($B_{eff}^{1/4}=136 MeV$) undergoes scalarization shows that $\beta=-4.5$ lies well below the critical coupling for all three bag constants. Stars with the stiffest EoS ($B_{eff}^{1/4}=128 MeV$) consistently scalarize at the lowest central densities and with the greatest strength, whereas those with the softest EoS ($B_{eff}^{1/4}=136 MeV$) require a more negative $\beta$ to achieve a comparable degree of scalarization. This behavior reflects the compactness-driven nature of the tachyonic instability: a more negative $\beta$ makes the effective mass of the scalar field more negative, so a lower compactness is sufficient to trigger scalarization. For the strongest coupling ($\beta=-6$), the effect of $B_{eff}$ on the mass of scalarized stars is more considerable. We can also probe the influences of the perturbative QCD contribution, $a_4$, in Normal model using Figure \ref{MroNormal}.
A larger $a_4$ (i.e., weaker perturbative QCD corrections) stiffens the SQS EoS, raising the maximum mass and producing a more compact star. That increased compactness deepens the gravitational potential and strengthens the tachyonic instability. Increasing $a_4$ slightly lowers the onset density for scalarization and yields a higher peak mass.
Our results confirm that, for the stiffest EoS ($a_4=0.7$), the deviation from the GR prediction is more pronounced.
Nevertheless, for $\beta=-6$, even the softest model ($a_4=0.5$) scalarizes strongly over an extremely wide interval.
Therefore for all three cases of $a_4$, $\beta=-6$ makes the scalarized branch dominant across the entire astrophysically relevant density range.
Our calculations verify that all three $a_4$ values produce scalarization, confirming that $\beta=-4.5$ lies below the critical coupling for scalarization. The sensitivity to $a_4$ is, however, milder than the sensitivity to the bag constant, which is consistent with the
EoS trends in Figure \ref{pNormal}.

In Figure \ref{MroCFL}, we present the mass of SQSs as a function of central density in the CFL model for strange quark matter.
In the CFL model, the key physical parameter is the uncertain CFL pairing gap, $\Delta$.
Due to the term $-\Delta^2\mu^2$ in the condensation energy, an extra repulsive contribution to the pressure is supplied and larger $\Delta$ stiffens the SQS EoS (see Figure \ref{pCFL}).  This stiffening increases the stellar mass, makes the star more compact, deepening the gravitational potential and strengthening the tachyonic instability. Consequently, the extra repulsion arising from the CFL pairing gap amplifies scalarization. Additionally, higher values of $\Delta$ cause scalarized stars to deviate from their GR counterparts at lower central densities, producing more pronounced scalarization effect.
Thus the uncertain CFL pairing gap, $\Delta$, acts as a scalarization enhancer, similar to the role of $a_4$ in the Normal model.
The existence of scalarization at $\beta=-4.5$ in the CFL model demonstrates that this coupling lies below the critical value for these CFL EoSs.
The differences among the three values of $\Delta$ are more significant at the strongest coupling considered ($\beta=-6$).
Comparing the CFL model with the Normal one shows that, for the SQS EoSs employed in this work, CFL stars are systematically less scalarized than those built with the stiffest Normal EoSs. This is because our stiffest CFL EoS ($\Delta=50 MeV$) is softer than the stiffest Normal EoS ($B_{eff}^{1/4}=128 MeV$).

Figure \ref{MroCFLm} shows the mass of SQSs versus the central density in the CFLm model. A larger $m_s$ softens the SQS EoS, reducing stellar compactness and thereby weakening the tachyonic instability. Consequently, scalarization is most pronounced for the stiffest EoS ($m_s=20 MeV$). For SQSs with $m_s=20 MeV$, the scalarized curve deviates from the GR one, exhibiting the earliest onset of scalarization and the largest mass enhancement. In contrast, for higher values of $m_s$, SQSs become scalarized at higher central densities with a smaller mass enhancement. These trends confirm that the strange quark mass acts as a scalarization suppressor, much like $B_{eff}$ in the Normal model. This behavior follows directly from Figure \ref{pCFLm}: a larger $m_s$ softens the EoS. Our results verify that, for all three values of $m_s$ considered, scalarization occurs at $\beta=-4.5$, demonstrating that this coupling lies below the critical value for the CFLm model. The influence of the strange quark mass on scalarized SQSs is more considerable for $\beta=-6$. The three figures demonstrate that the mass-density signature of spontaneous scalarization is universal across different descriptions of quark matter, with the coupling $\beta$ as the dominant control parameter. The stiffness of each quark matter model, controlled by $B_{eff}$, $a_4$, $\Delta$, and $m_s$, dictates the compactness of the star. Stiffer EoSs (smaller $B_{eff}$ or $m_s$, larger $a_4$ or $\Delta$) produce more compact
 stars, which deepen the gravitational potential and strengthen the tachyonic instability.

\begin{figure}[h]
	\subfigure{}{\includegraphics[scale=0.36]{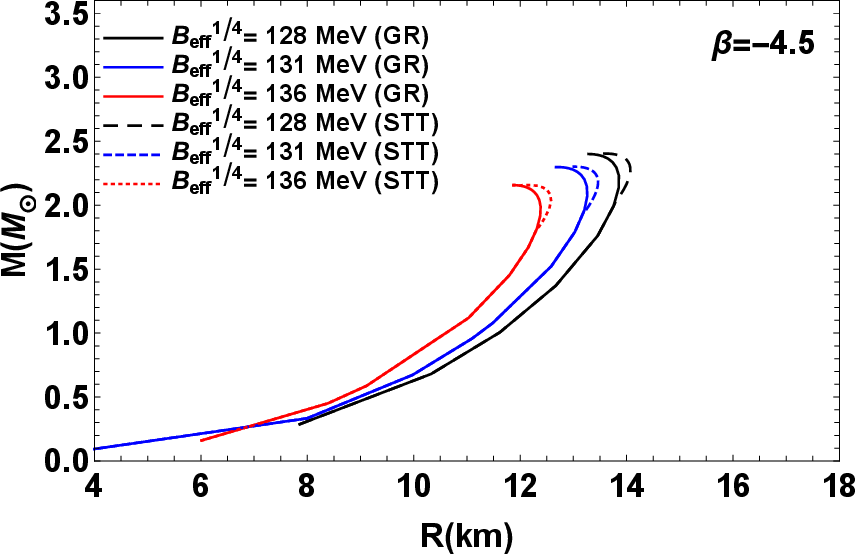}
		}	
\subfigure{}{\includegraphics[scale=0.36]{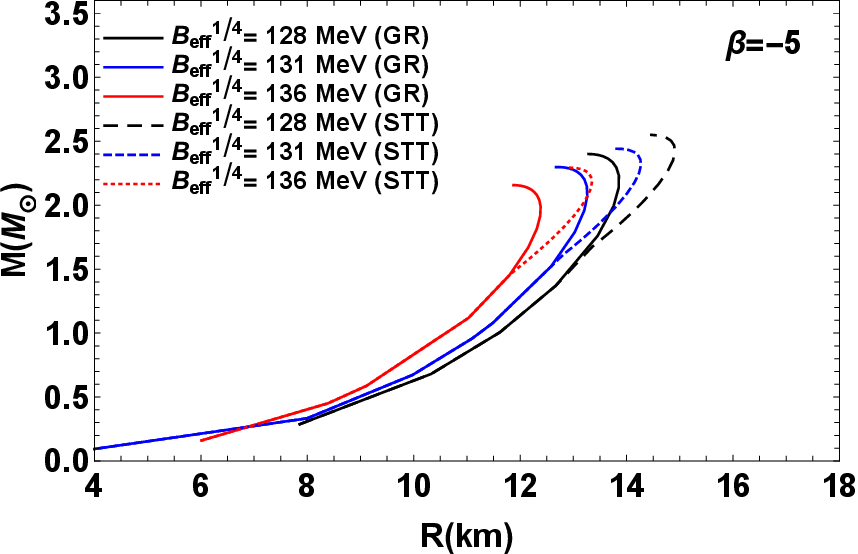}
		}
\subfigure{}{\includegraphics[scale=0.36]{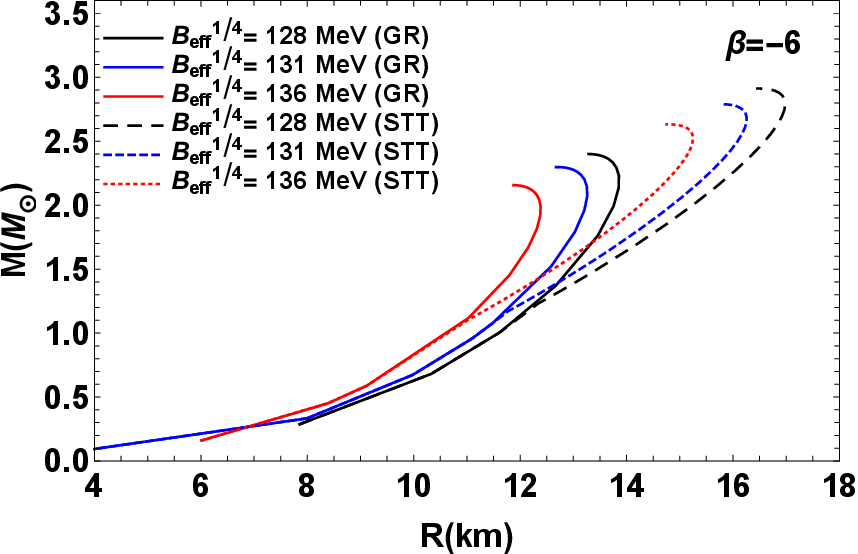}
		}\\
	\subfigure{}{\includegraphics[scale=0.36]{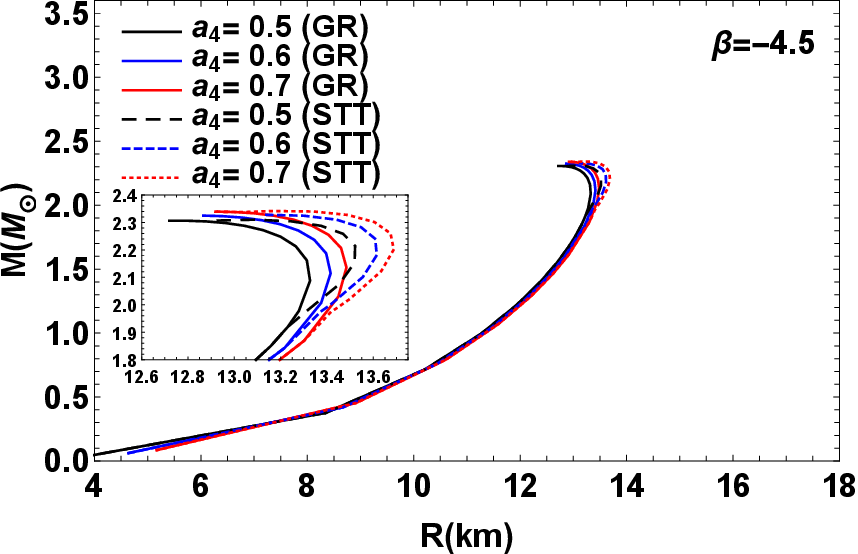}
		}	
\subfigure{}{\includegraphics[scale=0.36]{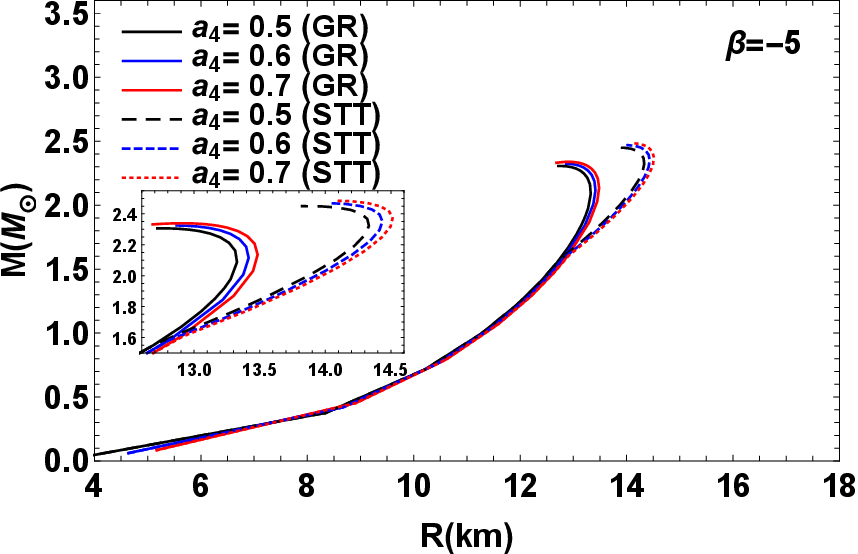}
		}
\subfigure{}{\includegraphics[scale=0.36]{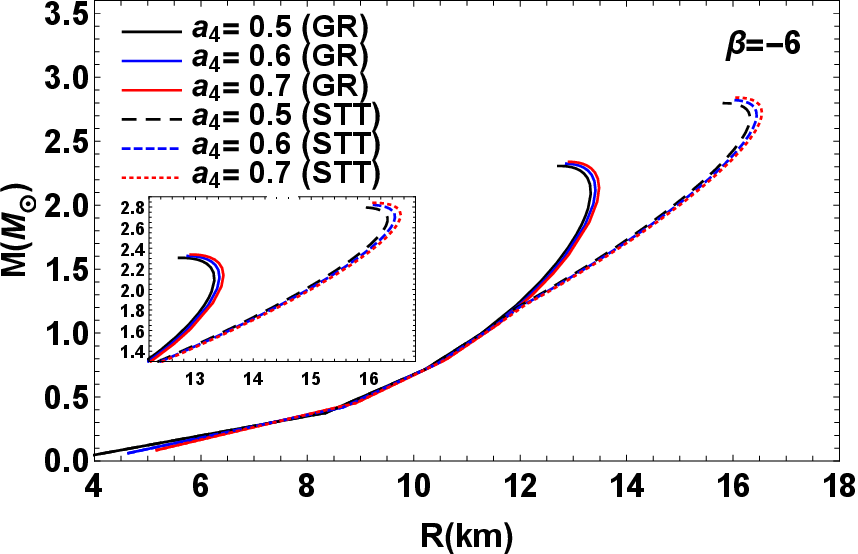}
		}
	\caption{Strange quark star mass versus the star radius in Normal model for a fixed strange quark mass $m_s=100MeV$, Top: considering the value of $a_4=0.58$ and different values of $B_{eff}$, and Bottom: considering the value of $B_{eff}^{1/4}=130.3MeV$ and different values of $a_4$, for three values of the coupling constant, $\beta$, in STT and GR. }
	\label{MrNormal}
\end{figure}

\begin{figure}[h]
	\subfigure{}{\includegraphics[scale=0.36]{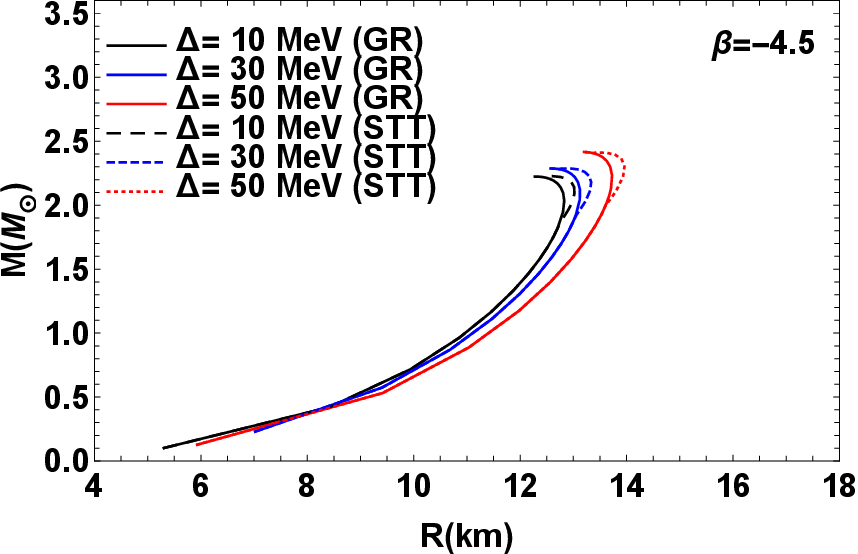}
		}	
	\subfigure{}{\includegraphics[scale=0.36]{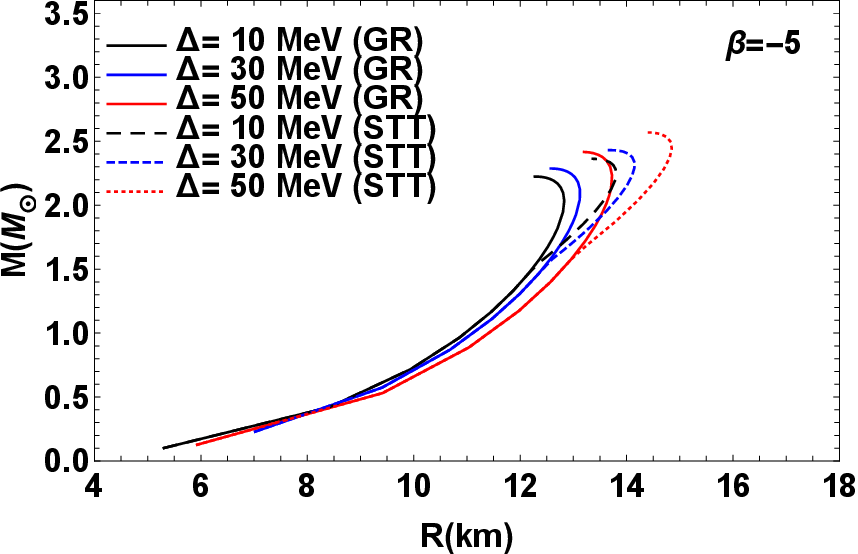}
		}
	\subfigure{}{\includegraphics[scale=0.36]{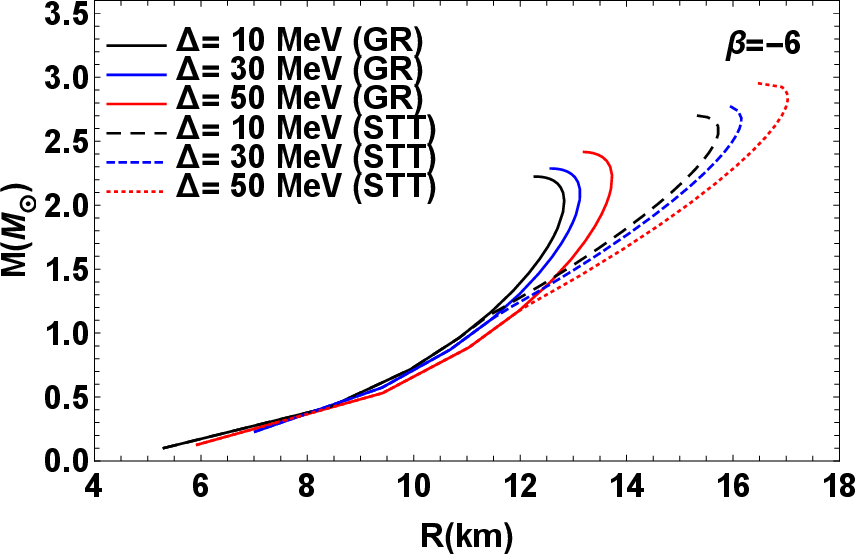}
		}
	\caption{Strange quark star mass versus the star radius considering CFL model with the values of $m_s=100MeV$, $a_4=0.53$, $B_{eff}^{1/4}=133.1MeV$, and different values of $\Delta$ for three values of the coupling constant, $\beta$, in STT and GR.}
	\label{MrCFL}
\end{figure}

\begin{figure}[h]
	\subfigure{}{\includegraphics[scale=0.36]{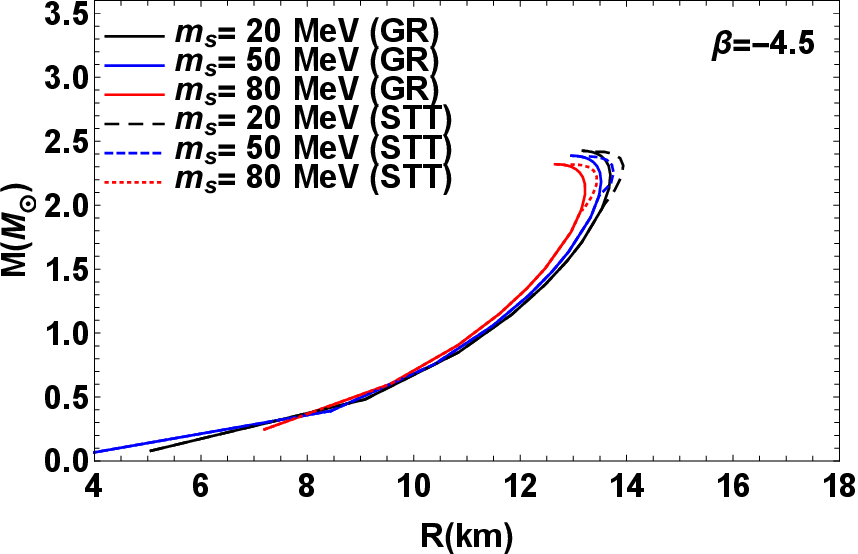}
		}	
	\subfigure{}{\includegraphics[scale=0.36]{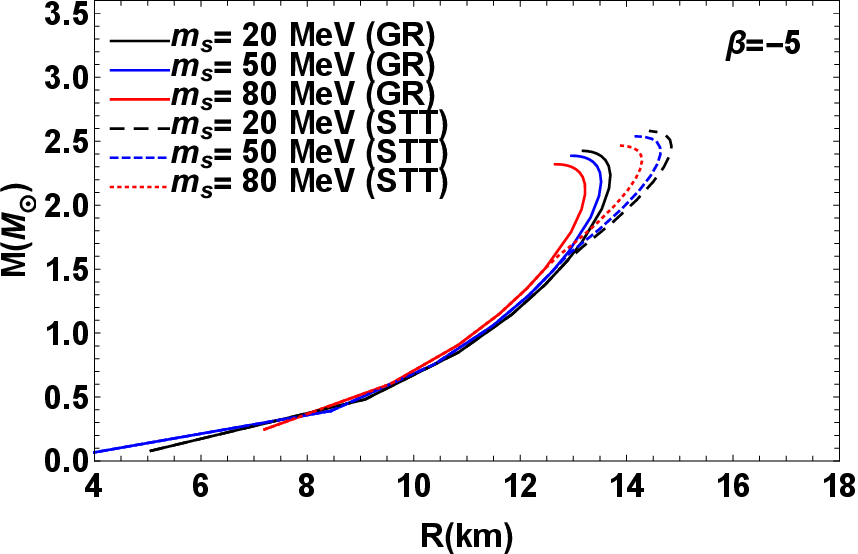}
		}
	\subfigure{}{\includegraphics[scale=0.36]{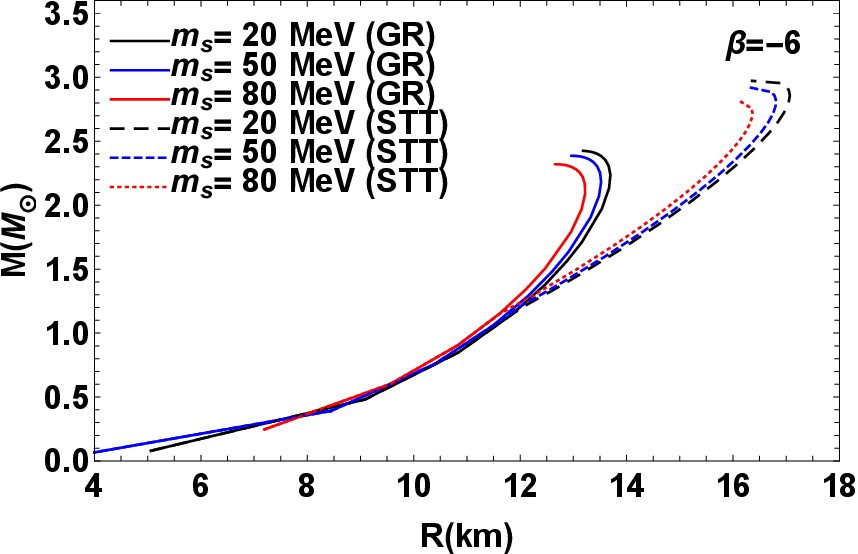}
		}
	\caption{Strange quark star mass versus the star radius in CFLm model with the values of $a_4=0.53$, $\Delta=33.9MeV$, $B_{eff}^{1/4}=134.4MeV$, and different values of $m_s$ for three values of the coupling constant, $\beta$, in STT and GR.}
	\label{MrCFLm}
\end{figure}

In Figures \ref{MrNormal}-\ref{MrCFLm}, we present the mass of SQSs versus their radius (mass-radius relation) for different models of strange quark matter. In the GR cases, the equilibrium sequences are purely gravitational: at low masses the stars are self-bound
and very small, and the radius increases to a maximum just before the mass peak. In scalar-tensor theory, when spontaneous scalarization occurs, the star develops additional pressure support, which modifies the mass-radius relation relative to the GR one. For scalarized stars, the maximum mass is significantly enhanced, especially for lower values of $\beta$. At a given mass, scalarized stars have larger radii than their GR counterparts, exhibiting a dramatic radius inflation for the strongest coupling considered ($\beta=-6$). For these stars, the mass-radius relation exhibits a deviation from the GR curve, corresponding to the scalarized branch shown in Figures \ref{MroNormal}-\ref{MroCFLm}. For more negative values of the coupling $\beta$, the deviation becomes more pronounced, consistent with the scalarized branch shown in the previous figures. For scalarized stars, as the central density increases, the stellar radius first increases and then decreases as the star approaches the maximum mass.

The mass-radius relation for Normal SQSs is depicted in Figure \ref{MrNormal}. The stiffest EoS (smallest $B_{eff}$) produces the strongest scalarization signature, the largest radius inflation and the highest maximum mass relative to GR case (see Figure \ref{pNormal}). For a softer EoS (i.e., larger $B_{eff}$), both the stellar mass and radius decrease, and the deviation in the mass-radius relation of scalarized stars becomes smaller. The scalarized branch in the mass-radius relation is most affected by the stiffest EoS ($B_{eff}^{1/4}=128 MeV$), which produces the most compact star and thus the strongest gravitational potential, while the softest EoS shows a smaller but still substantial deviation from the GR curve. The effects of the perturbative QCD contribution, $a_4,$ on the mass-radius relation of SQSs are also worth investigating. The scalarized branch is influenced by $a_4$: larger values of $a_4$ yield slightly larger radii and higher maximum masses. This effect directly follows Figure \ref{pNormal}, where larger $a_4$ stiffens the EoS, making the star more compact and more susceptible to the tachyonic instability.

The results for the mass-radius relation in the CFL model are presented in Figure \ref{MrCFL}. A larger CFL pairing gap, $\Delta$, stiffens the EoS (Figure \ref{pCFL}), making the star more compact and therefore more susceptible to the tachyonic instability.
For the stiffest EoS ($\Delta=50MeV$), the deviation is largest, with a clear radius excess and an enhanced maximum mass.
As the coupling becomes more negative, the sensitivity to the pairing gap increases, and the scalarization signature in the mass-radius plane becomes a sensitive probe of the CFL phase.

The mass-radius relation for SQSs in the CFLm model is shown in Figure \ref{MrCFLm}. All three values of the strange quark mass yield clear scalarized branches, with radii significantly larger than those of the corresponding GR curves and with enhanced maximum masses. The size and shape of these branches depend on $m_s$: the stiffest EoS ($m_s=20 MeV$) produces the largest deviation from the GR curve. In contrast, the softest one ($m_s=80 MeV$) produces a smaller but still robust branch. This is because a smaller $m_s$ stiffens the EoS (see Figure \ref{pCFLm}), increasing the stellar compactness and thereby
favoring the tachyonic instability. The CFLm stars are systematically softer than both the Normal and CFL models at the same coupling, so scalarization is even more suppressed in the CFLm model. We find from Figures \ref{MrNormal}-\ref{MrCFLm} that all three models of strange quark matter scalarize robustly, with clear mass-radius branches and a pronounced sensitivity to the microphysical parameter. Smaller $B_{eff}$, larger $a_4$, larger $\Delta$, and smaller $m_s$ enhance the scalarization signature. Therefore, spontaneous scalarization leaves a universal mass-radius signature across all three
 models: a scalarized branch whose size is controlled primarily by $\beta$ and fine-tuned by the specific microphysical stiffness of the EoS.

\subsection{Central Scalar field In Strange Quark Star}\label{s}

\begin{figure}[h]
	\subfigure{}{\includegraphics[scale=0.36]{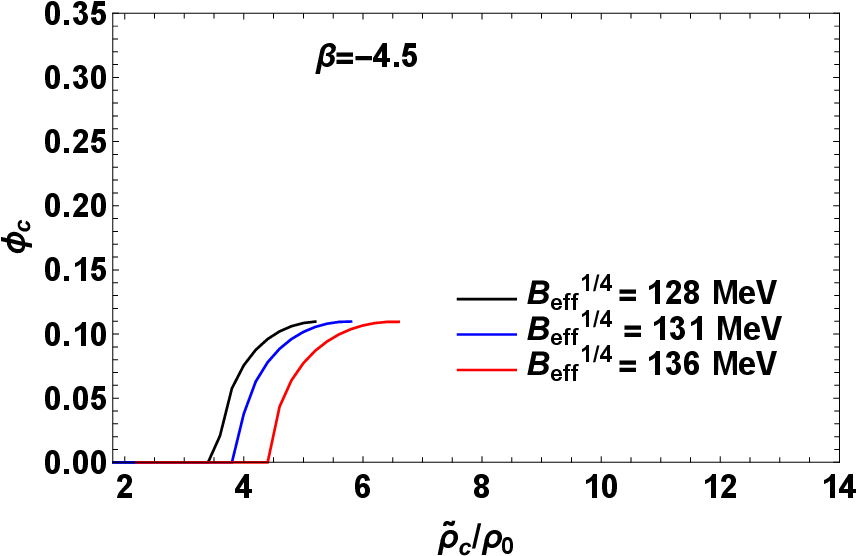}
		}	
\subfigure{}{\includegraphics[scale=0.36]{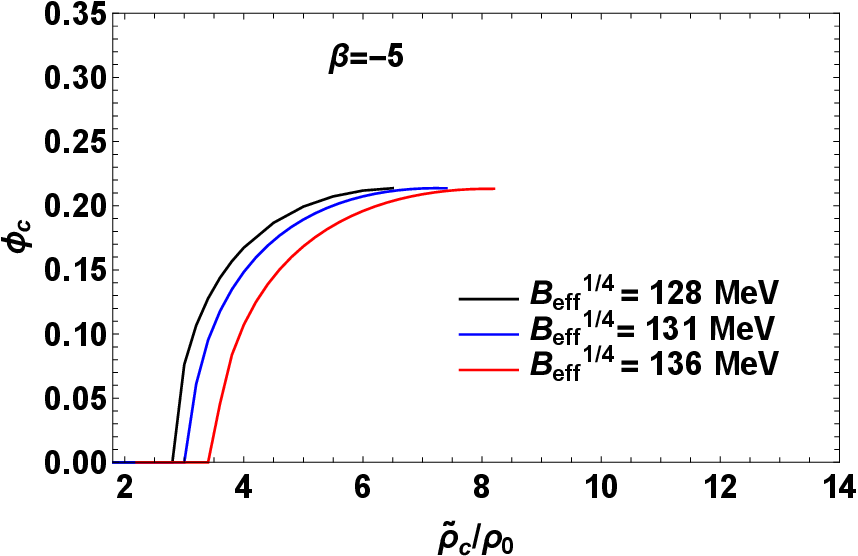}
		}
\subfigure{}{\includegraphics[scale=0.36]{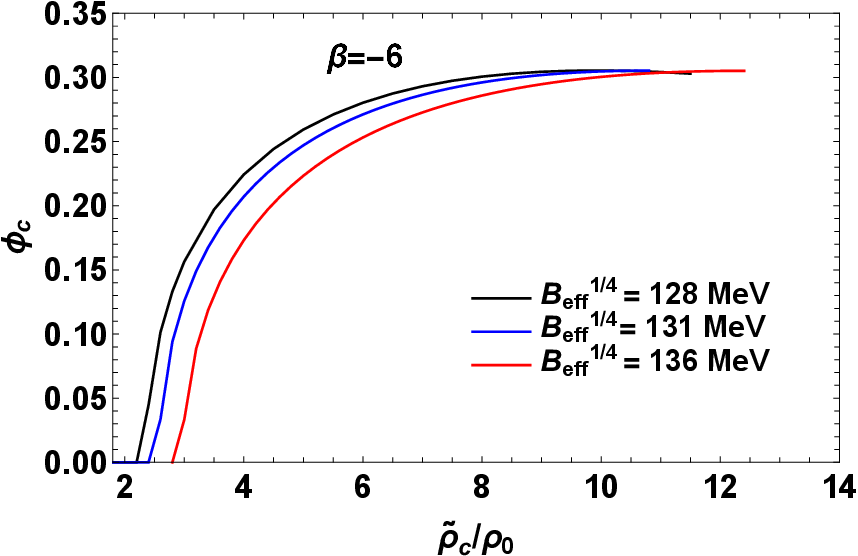}
		}\\
	\subfigure{}{\includegraphics[scale=0.36]{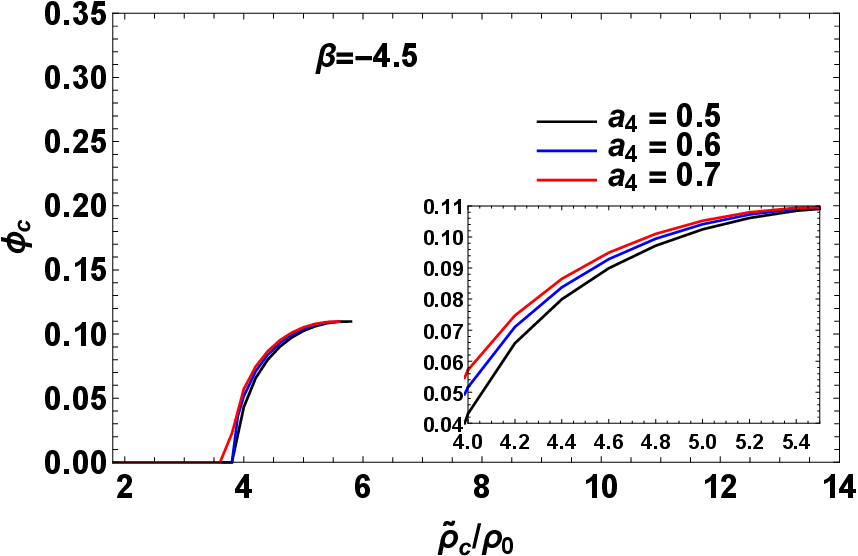}
		}	
\subfigure{}{\includegraphics[scale=0.36]{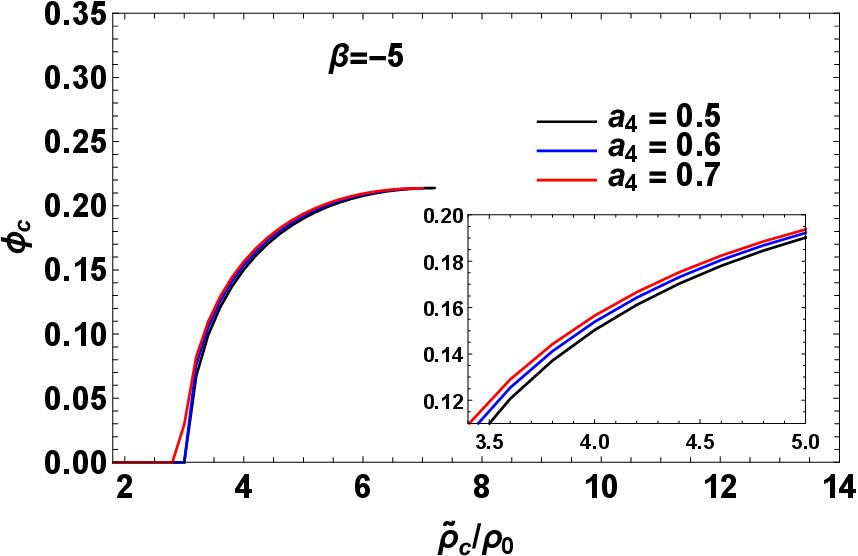}
		}
\subfigure{}{\includegraphics[scale=0.36]{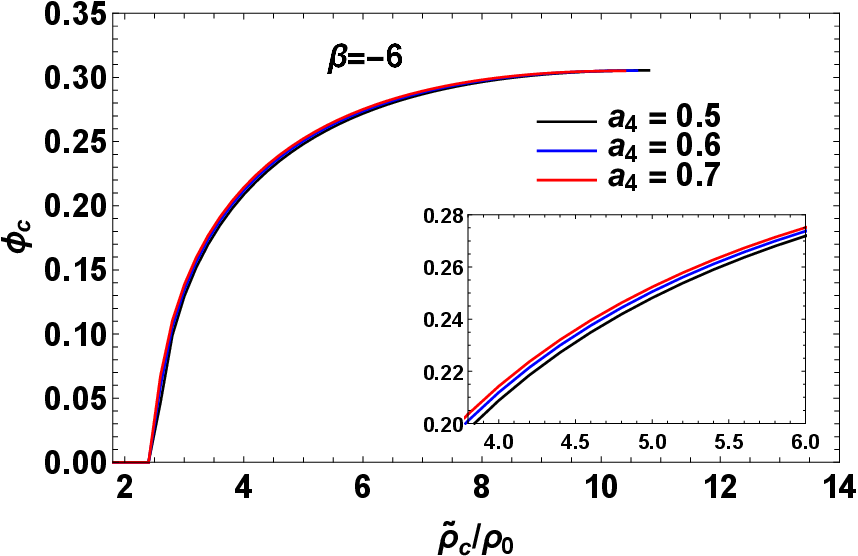}
		}
	\caption{Central scalar field versus the central density in Normal model with the value of $m_s=100MeV$, Top: considering the value of $a_4=0.58$ and different values of $B_{eff}$, and Bottom: considering the value of $B_{eff}^{1/4}=130.3MeV$ and different values of $a_4$, for three values of the coupling constant $\beta$.}
	\label{phiNormal}
\end{figure}

\begin{figure}[h]
	\subfigure{}{\includegraphics[scale=0.36]{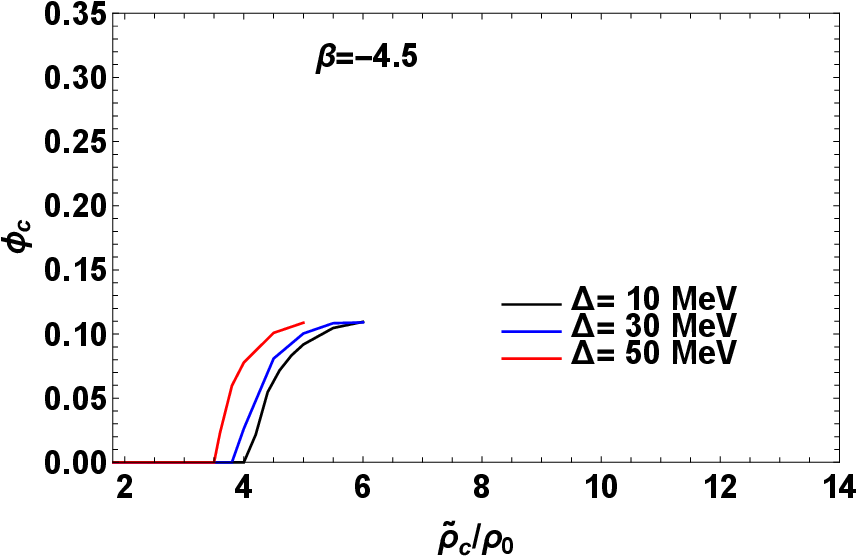}
		}	
	\subfigure{}{\includegraphics[scale=0.36]{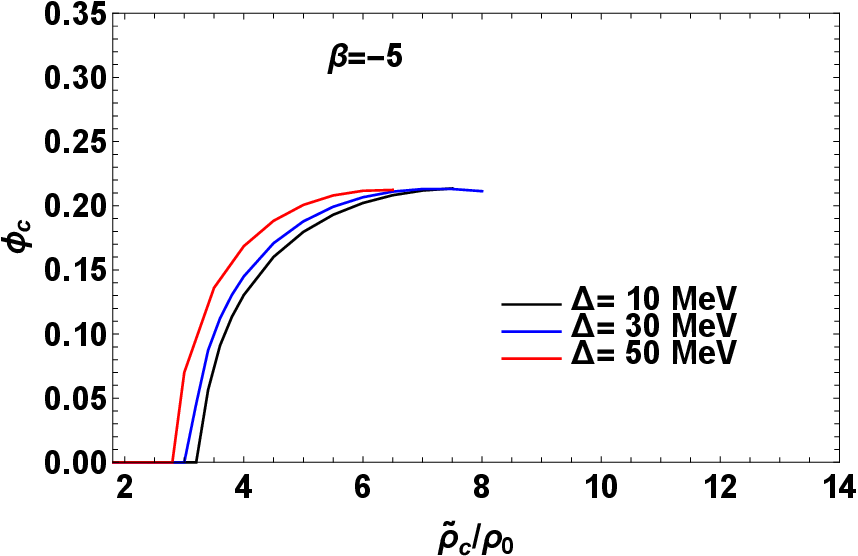}
		}
	\subfigure{}{\includegraphics[scale=0.36]{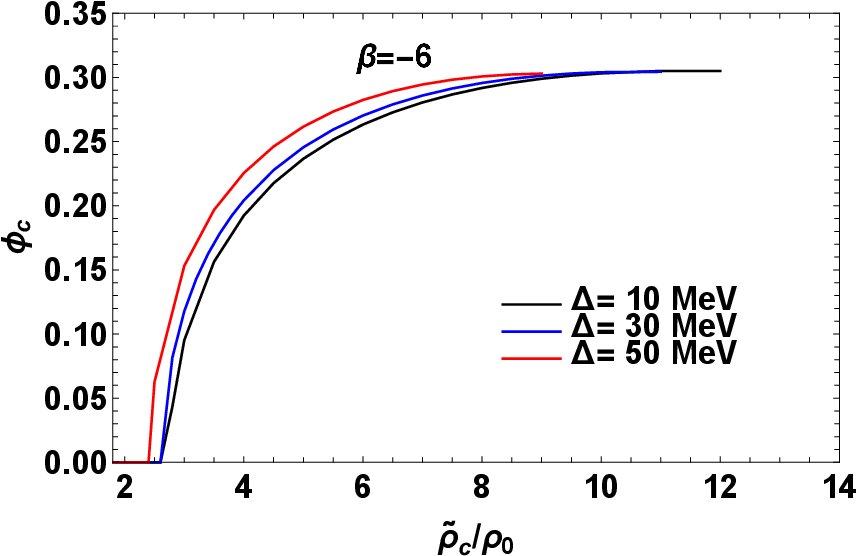}
		}
	\caption{Central scalar field versus the central density in CFL model with the values of $m_s=100MeV$, $a_4=0.53$, $B_{eff}^{1/4}=133.1MeV$, and different values of $\Delta$ for three values of the coupling constant, $\beta$.}
	\label{phiCFL}
\end{figure}

\begin{figure}[h]
	\subfigure{}{\includegraphics[scale=0.36]{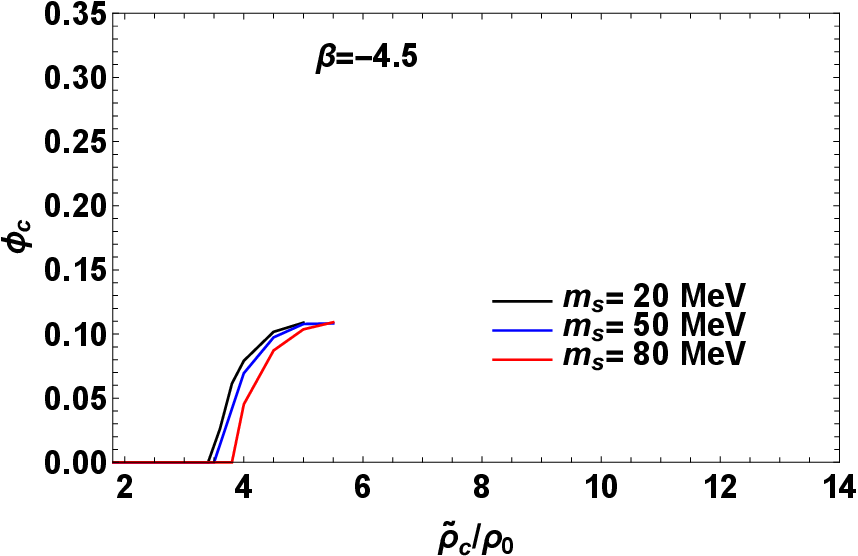}
		}	
	\subfigure{}{\includegraphics[scale=0.36]{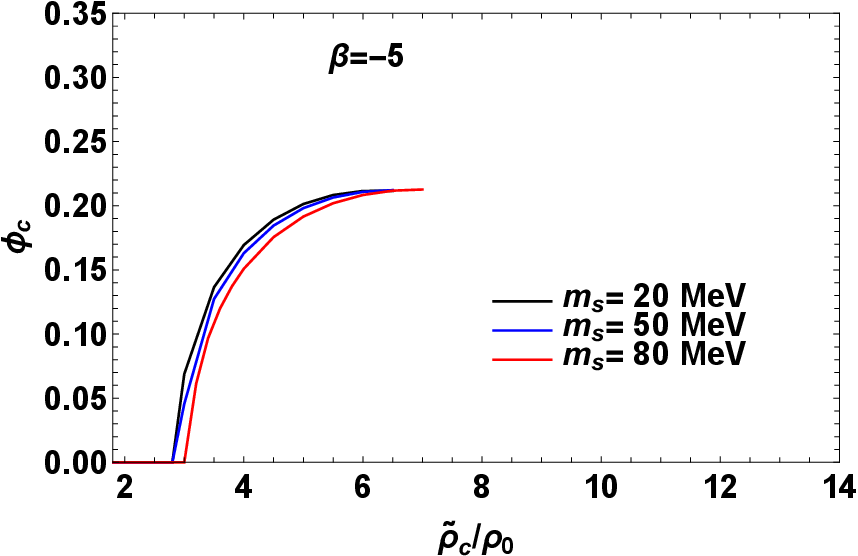}
		}
	\subfigure{}{\includegraphics[scale=0.36]{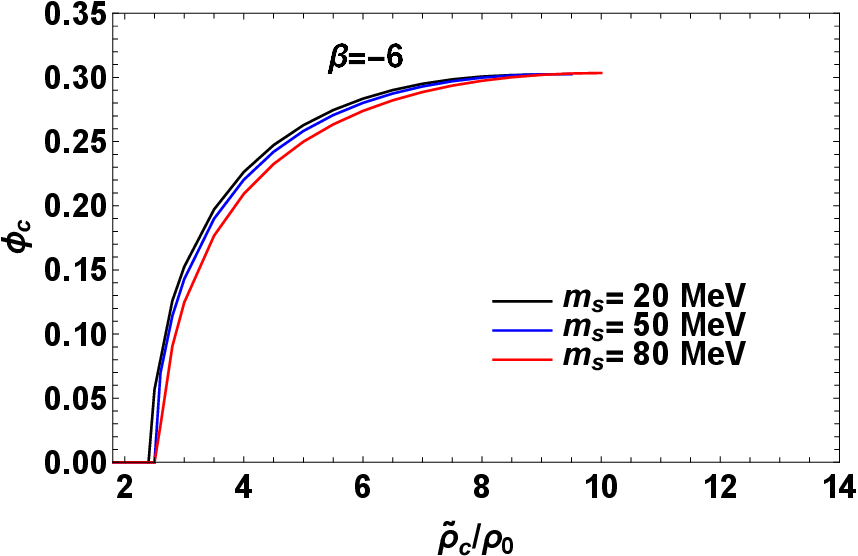}
		}
	\caption{Central scalar field versus the central density in CFLm model with the values of $a_4=0.53$, $\Delta=33.9MeV$, $B_{eff}^{1/4}=134.4MeV$, and different values of $m_s$ for three values of the coupling constant, $\beta$.}
	\label{phiCFLm}
\end{figure}

Figures \ref{phiNormal}-\ref{phiCFLm} show the central scalar field as a function of central density for SQSs in different models of strange quark matter. The central scalar field, $\phi_c$, is zero at low densities, rises from zero and reaches to a peak.
The central density at which $\phi_c$ becomes nonzero is referred to as the critical density, $\rho_{crit}$. For more negative values of $\beta$, the scalarization range (where $\phi_c\neq0$) widens considerably: the onset of scalarization shifts to lower central densities, the peak central scalar field increases, and the scalarized phase extends to much higher densities.
Our results indicate that for $\beta=-6$, scalarization persists deep into the high-density regime. At this coupling, the scalar field is a robust feature of the star, persisting across essentially the entire stable mass range.

Figure \ref{phiNormal} confirms that a larger bag constant (softer EoS) shifts the onset of scalarization ($\rho_{crit}$) to higher central densities, because a less compact star needs a higher density to reach sufficient compactness for the tachyonic instability. The peak value of $\phi_c$ is nearly the same for all three values of both $B_{eff}$ and $a_4$, indicating that at a given coupling the maximum central scalar field strength is largely independent of the EoS stiffness. The scalarization range (the interval of central densities where $\phi_c\neq0$) grows with $B_{eff}$, consistent with the broader scalarized branches in Figures \ref{MroNormal} and \ref{MrNormal}. Moreover, a larger $a_4$ (stiffer EoS) shifts the onset of scalarization to lower densities.

In Figure \ref{phiCFL}, we present the central scalar field versus the central density in the CFL model of strange quark matter. A larger $\Delta$ (stiffer EoS) shifts the entire range of scalarization to lower densities and $\rho_{crit}$ decreases. This is the compactness effect which a stiffer EoS makes the star more compact at a given central density, so it reaches the critical compactness earlier (lower onset) because the star reaches its maximum mass at a lower central density. The stiffest EoS ($\Delta=50MeV$) has the narrowest scalarized interval, while the softest one ($\Delta=10MeV$) scalarizes over a broader density range. Moreover, the peak value of $\phi_c$ is nearly identical for all three values of $\Delta$.

Figure \ref{phiCFLm} displays the central scalar field in the CFLm model. The smaller $m_s$ (stiffer EoS) moves the onset $\rho_{crit}$ to lower densities because the more compact star reaches the critical compactness earlier. However, the peak central scalar field $\phi_c$ remains nearly unchanged as $m_s$ increases. The scalarization range also narrows as $m_s$ decreases. These $\phi_c$ curves (Figures \ref{phiNormal}-\ref{phiCFLm}) directly generate the mass-density deviations shown in Figures \ref{MroNormal}-\ref{MroCFLm} and the scalarized mass-radius branches displayed in Figures \ref{MrNormal}-\ref{MrCFLm}, confirming that the compactness-driven scalarization mechanism is universal across the three descriptions of strange quark matter.

\begin{figure}[h]
	\subfigure{}{\includegraphics[scale=0.5]{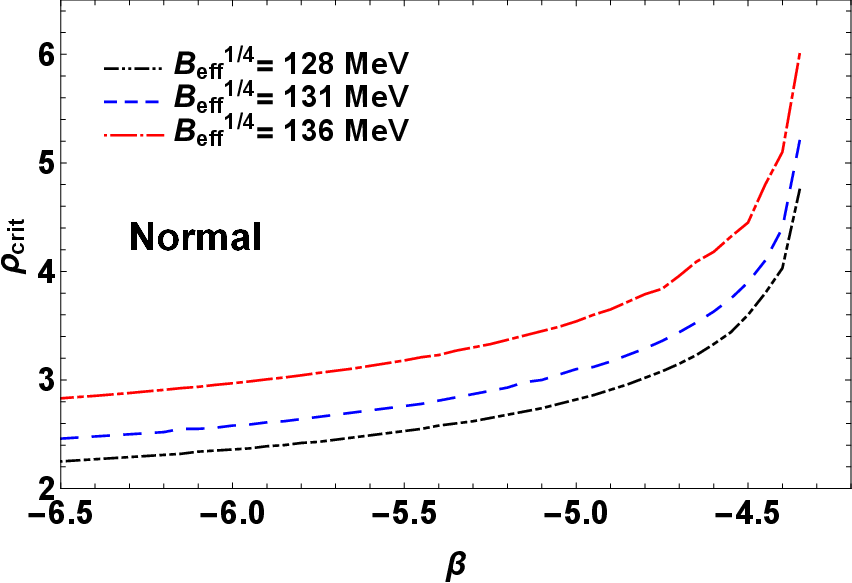}
		}	
	\subfigure{}{\includegraphics[scale=0.5]{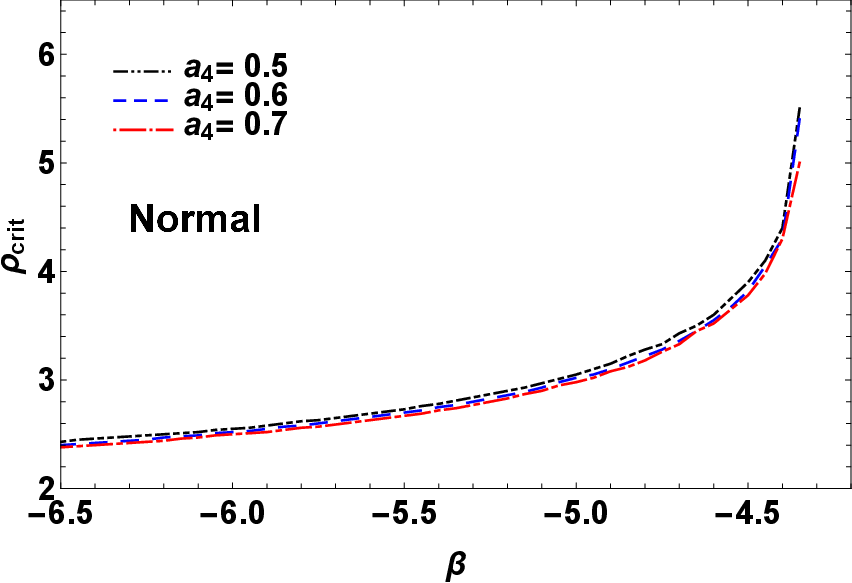}
		}	\\
\subfigure{}{\includegraphics[scale=0.5]{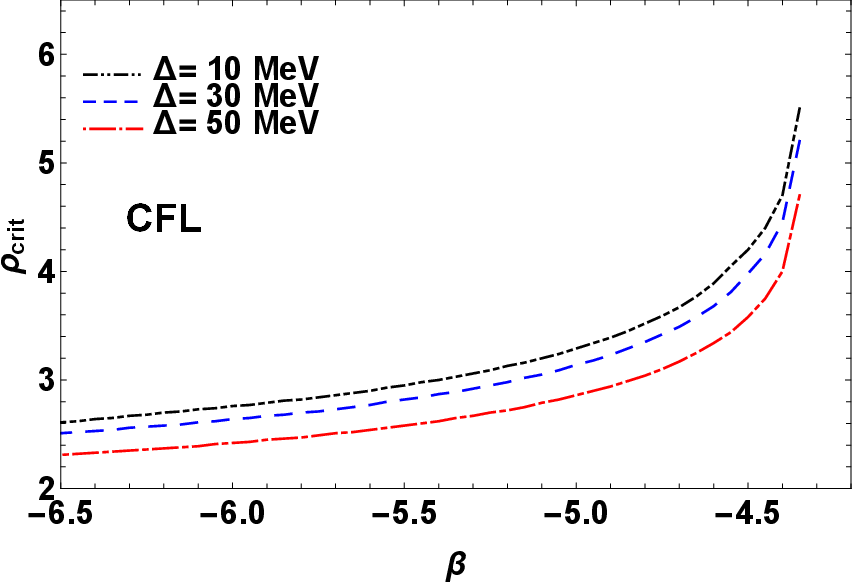}
		}	
\subfigure{}{\includegraphics[scale=0.5]{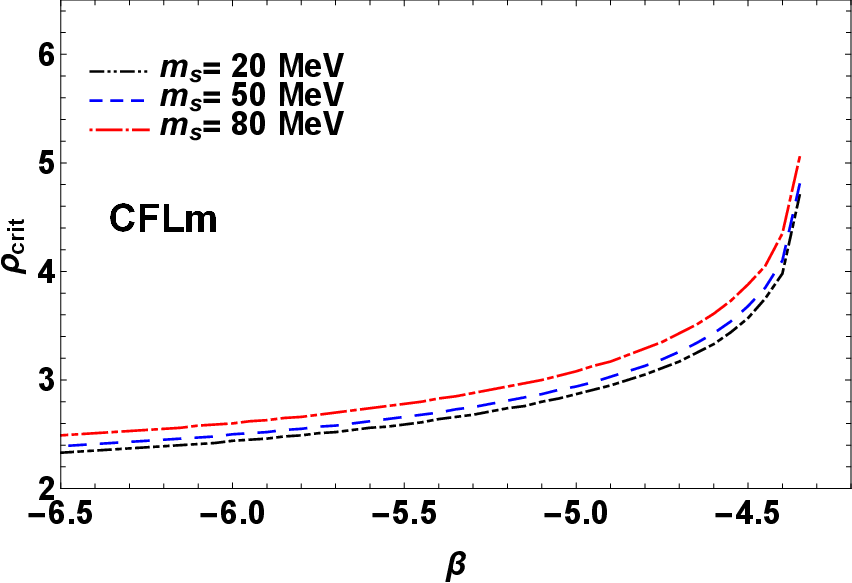}
		}
	\caption{Critical density of scalarization, $\rho_{crit}$, versus the coupling constant, $\beta$, in Normal, CFL, and CFLm models.
For Normal model, we have considered the value $m_s=100MeV$, as well as (left): $a_4=0.58$ and different values of $B_{eff}$ and (right): $B_{eff}^{1/4}=130.3MeV$ and different values of $a_4$. For CFL model, we have assumed the values of $m_s=100MeV$, $a_4=0.53$, $B_{eff}^{1/4}=133.1MeV$, and different values of $\Delta$. Besides, for CFLm model, we have considered the values of $a_4=0.53$, $\Delta=33.9MeV$, $B_{eff}^{1/4}=134.4MeV$, and different values of $m_s$.}
	\label{Firstcd}
\end{figure}

The onset density of scalarization for SQSs is shown in Figure \ref{Firstcd}. This figure supports the compactness-driven picture: the coupling strength and the EoS stiffness jointly control the onset density of scalarization. Figure \ref{Firstcd} shows that a more negative coupling constant, $\beta$, (i.e., stronger coupling) reduces the onset density $\rho_{crit}$, thereby shifting scalarization to lower central densities. The critical density grows as the coupling constant, $\beta$, increases (i.e., becomes less negative) because the strength of the tachyonic instability diminishes. In fact, the coupling becomes too weak to destabilize the scalar field unless the star is extremely compact, so the onset density rises sharply and eventually diverges as the coupling approaches the critical value. In the Normal model of SQS, at fixed $\beta$, a larger bag constant (softer EoS) yields a higher $\rho_{crit}$, because a less compact star requires a larger density to trigger the tachyonic instability. In this model, the curves remain nearly parallel, indicating that the effect of $B_{eff}$ on the onset density remains systematic across the whole coupling range. For a given value of $\beta$, a larger $a_4$ (stiffer EoS) leads to a slightly lower $\rho_{crit}$, because a stiffer star is more compact and therefore reaches the tachyonic instability threshold at a lower central density. The curves for different $a_4$ lie very close together, showing that the perturbative QCD parameter has a weaker influence on the onset density than the bag constant, $B_{eff}$. In the CFL model, a larger pairing gap $\Delta$ (stiffer EoS) reduces $\rho_{crit}$, because increased compactness triggers the tachyonic instability at a lower central density. This trend is the opposite of that for the bag constant, $B_{eff}$, where a softer EoS increased $\rho_{crit}$. Therefore, any parameter that stiffens the EoS (larger $\Delta$ or smaller $B_{eff}$) shifts the scalarization onset to lower densities, whereas a softening parameter has the opposite effect, pushing the onset to higher densities. The curves corresponding to the three values of $\Delta$ are nearly parallel, demonstrating that the effect of $\Delta$ on the onset density is systematic across the entire coupling range. In the CFLm model, a smaller $m_s$ (stiffer EoS) gives a smaller $\rho_{crit}$ because the more compact star reaches the critical compactness sooner. These curves are also nearly parallel, showing that the strange quark mass systematically shifts the onset density.

\subsection{Scalar Charge In Strange Quark Star}\label{s}

The scalar charge provides a more direct discriminator between GR and scalar-tensor theory than the mass-radius relation alone. In the GR limit, $\beta=0$, the scalar field vanishes identically and therefore $\omega=0$ for every stellar configuration. In contrast, scalarized SQSs possess a nonzero scalar charge, $\omega\neq0$. Thus, even when the mass and radius of a scalarized SQS are close to their GR counterparts, the presence of a nonzero scalar charge provides an unambiguous theoretical signature of the scalar degree of freedom.

\begin{figure}[h]
	\subfigure{}{\includegraphics[scale=0.36]{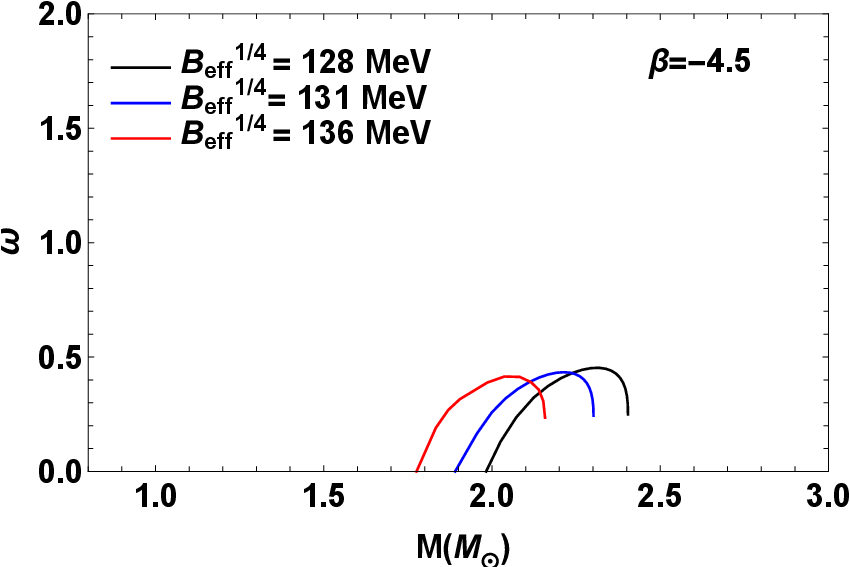}
		}	
\subfigure{}{\includegraphics[scale=0.36]{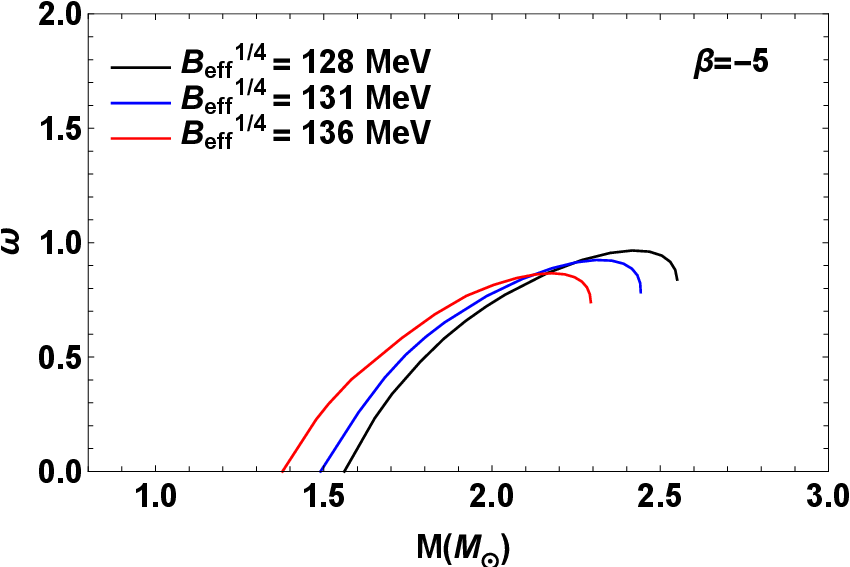}
		}
\subfigure{}{\includegraphics[scale=0.36]{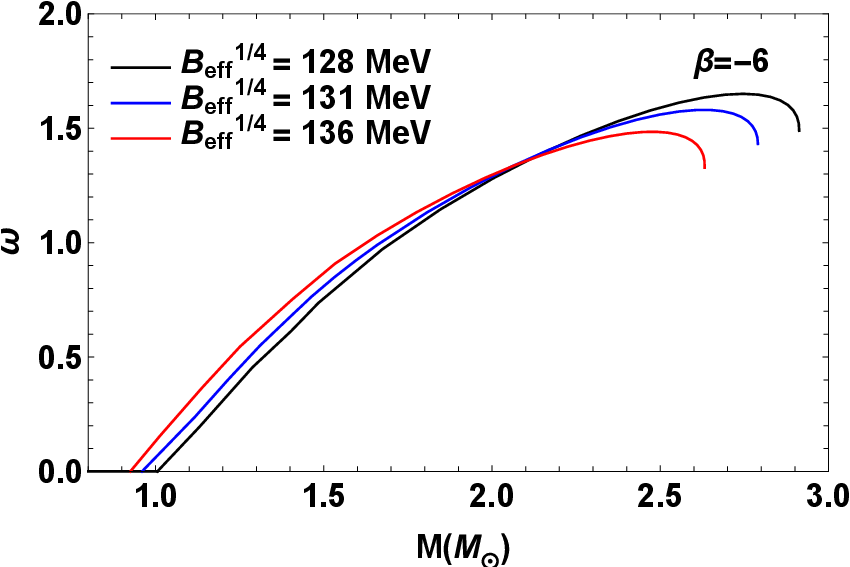}
		}\\
	\subfigure{}{\includegraphics[scale=0.36]{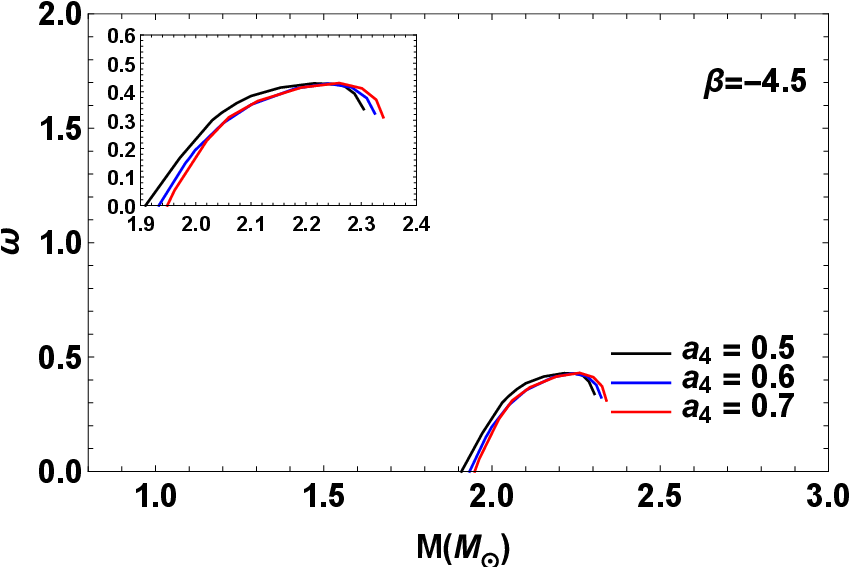}
		}	
\subfigure{}{\includegraphics[scale=0.36]{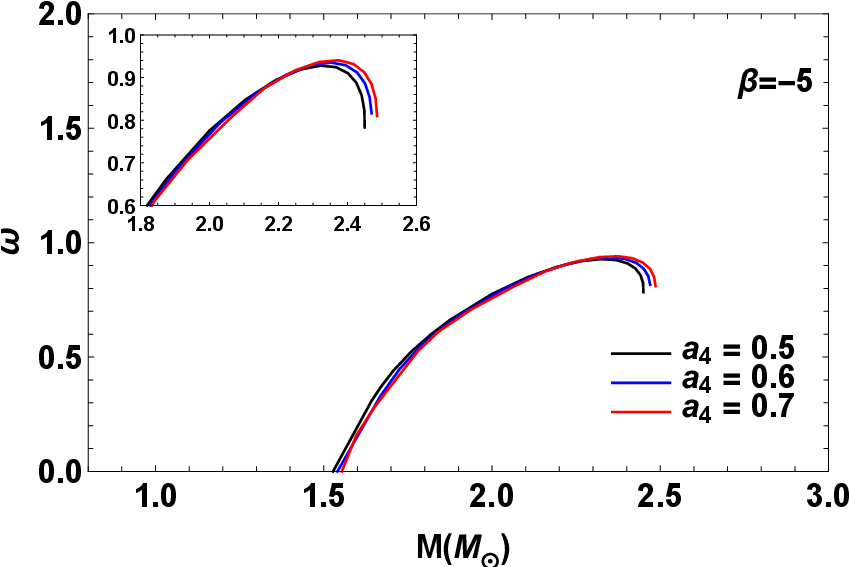}
		}
\subfigure{}{\includegraphics[scale=0.36]{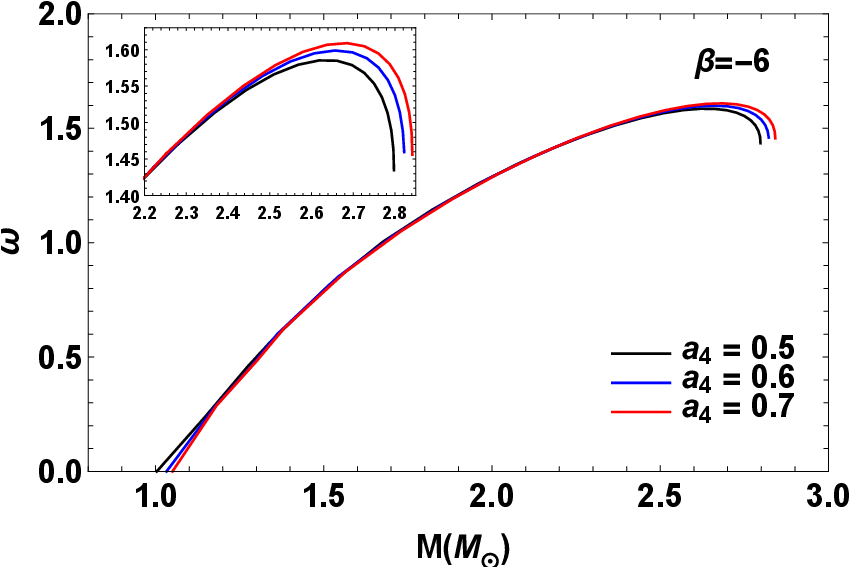}
		}
	\caption{Scalar charge, $\omega$, versus the star mass, $M$, in Normal model with the value of $m_s=100MeV$, Top: considering the value of $a_4=0.58$ and different values of $B_{eff}$, and Bottom: considering the value of $B_{eff}^{1/4}=130.3MeV$ and different values of $a_4$, for three values of the coupling constant $\beta$.}
	\label{omNormal}
\end{figure}

\begin{figure}[h]
	\subfigure{}{\includegraphics[scale=0.36]{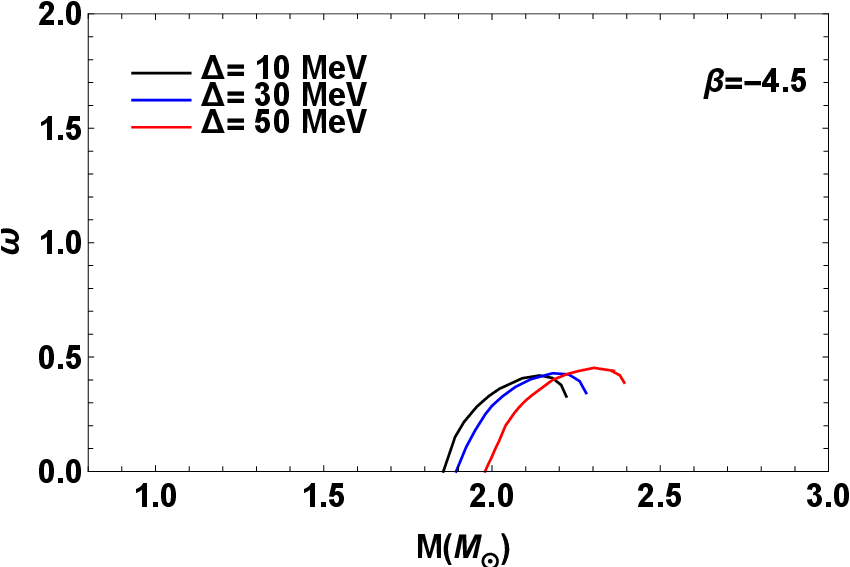}
		}	
	\subfigure{}{\includegraphics[scale=0.36]{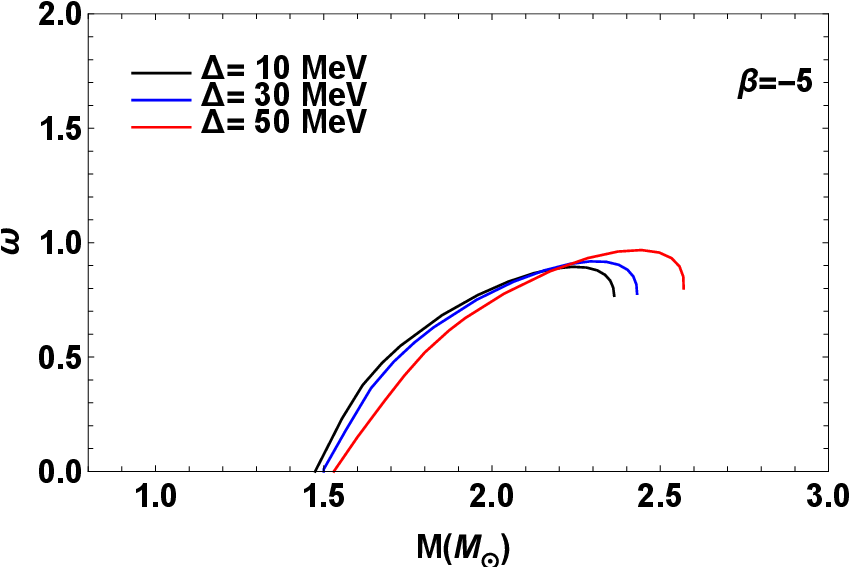}
		}
	\subfigure{}{\includegraphics[scale=0.36]{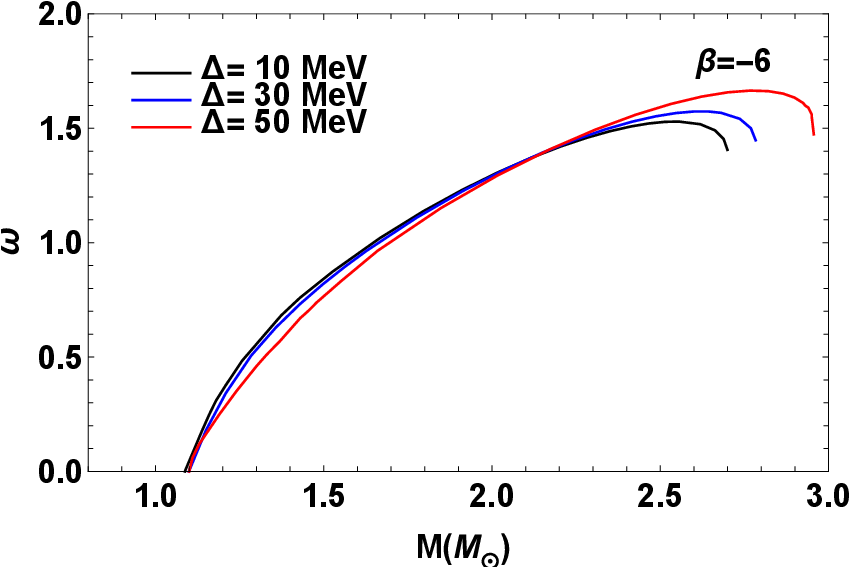}
		}
	\caption{Scalar charge, $\omega$, versus the star mass, $M$, in CFL model with the values of $m_s=100MeV$, $a_4=0.53$, $B_{eff}^{1/4}=133.1MeV$, and different values of $\Delta$ for three values of the coupling constant, $\beta$.}
	\label{omCFL}
\end{figure}

\begin{figure}[h]
	\subfigure{}{\includegraphics[scale=0.36]{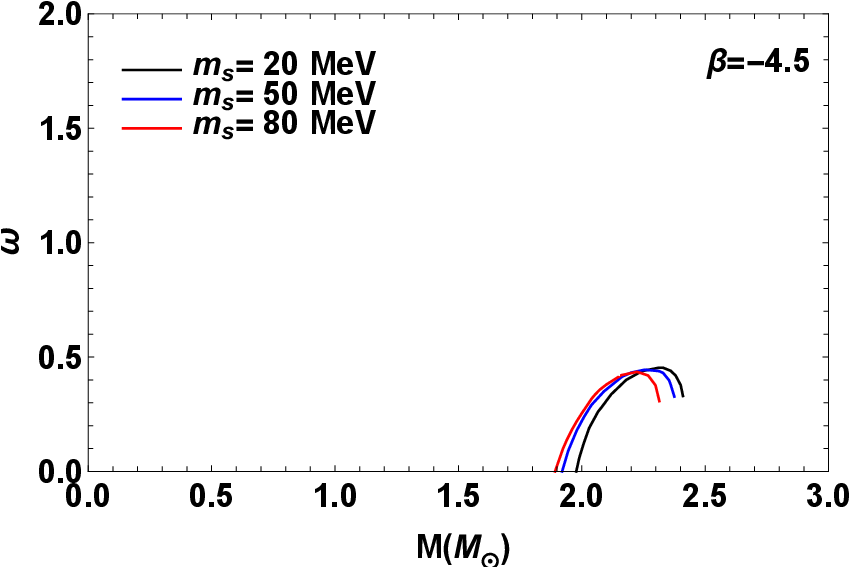}
		}	
	\subfigure{}{\includegraphics[scale=0.36]{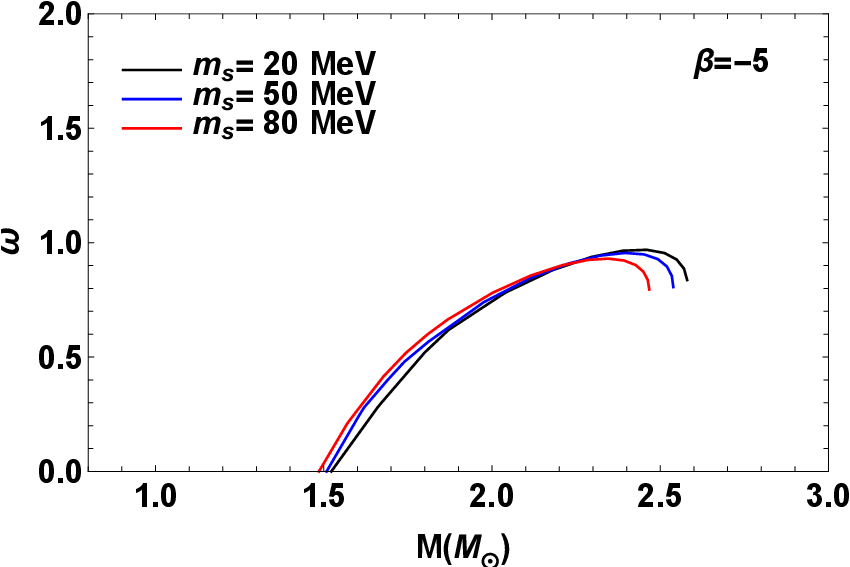}
		}
	\subfigure{}{\includegraphics[scale=0.36]{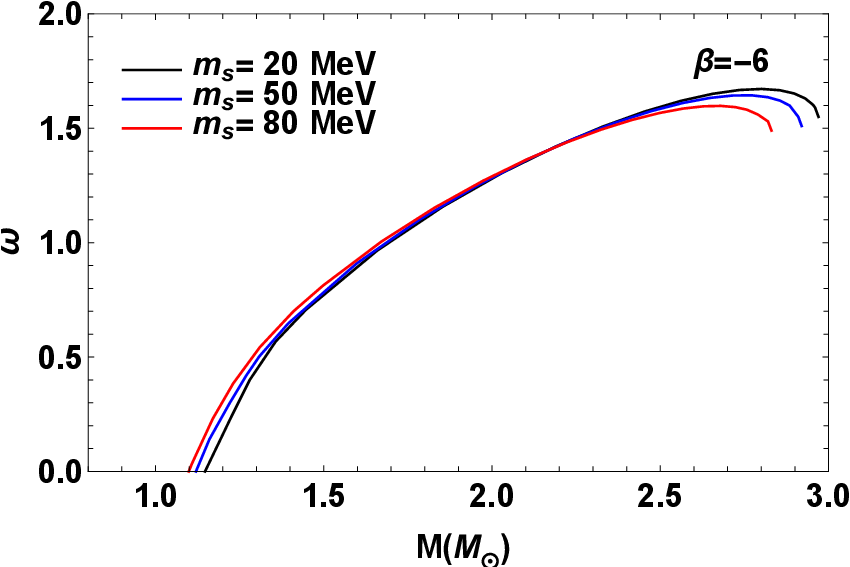}
		}
	\caption{Scalar charge, $\omega$, versus the star mass, $M$, in CFLm model with the values of $a_4=0.53$, $\Delta=33.9MeV$, $B_{eff}^{1/4}=134.4MeV$, and different values of $m_s$ for three values of the coupling constant, $\beta$.}
	\label{omCFLm}
\end{figure}

Figures \ref{omNormal}-\ref{omCFLm} exhibit the scalar charge, $\omega$, of SQSs as a function of stellar mass for different models of strange quark matter.
In each curve, $\omega$ is zero for low-mass unscalarized stars, rises to a peak, and then decreases slightly at higher masses. The nonzero $\omega$ interval exactly matches the scalarization extent seen in the mass-density (Figures \ref{MroNormal}-\ref{MroCFLm}), mass-radius (Figures \ref{MrNormal}-\ref{MrCFLm}), and central scalar field (Figures \ref{phiNormal}-\ref{phiCFLm}) figures, providing the observable imprint of spontaneous scalarization that would govern dipole radiation in binary systems. At stronger (more negative) coupling, $\omega$ reaches a much larger amplitude and spans a wider mass range, consistent with the robust scalarization observed in the mass-density, mass-radius, and central scalar field figures. Moreover, as $\beta$ decreases (i.e., becomes more negative), the SQS mass at which the scalar charge becomes nonzero drops to much lower values, while the maximum mass of the scalarized branch increases significantly. The mass intervals of nonzero scalar charge in Figures \ref{omNormal}-\ref{omCFLm} coincide exactly with the scalarization range shown in Figures \ref{phiNormal}-\ref{phiCFLm}. In addition, the scalar charge amplitude directly tracks the central scalar field, $\phi_c$, displayed in those figures.

Figure \ref{omNormal} confirms that in the Normal model, a stiffer EoS (smaller $B_{eff}$) leads to a larger maximum $\omega$ and a broader mass range of nonzero scalar charge, because the greater compactness drives a stronger scalar field and supports a higher maximum mass.
Increasing $B_{eff}$ shifts the scalar charge range to lower masses and reduces the peak scalar charge. Moreover, in the Normal model, a larger $a_4$ (stiffer EoS) results in a slightly larger maximum $\omega$ and a slightly wider mass range of nonzero scalar charge, because the enhanced compactness drives a stronger scalar field. However, the differences among the curves for different values of $a_4$ are modest, owing to the weaker influence of $a_4$ compared to that of $B_{eff}$.

In the CFL model (Figure \ref{omCFL}), a larger pairing gap, $\Delta$, (stiffer EoS) leads to a larger peak $\omega$ and a wider mass range, because the extra stiffness makes the star more compact and strengthens the scalar field. The SQS mass at which the scalar charge becomes nonzero increases with $\Delta$ especially for higher $\beta$ (weaker coupling). Moreover, the maximum mass that supports a nonzero scalar charge is larger for larger $\Delta$. Figure \ref{omCFLm} verifies that, for smaller $m_s$ (stiffer EoS), the scalar charge $\omega$ reaches a larger maximum and exhibits a wider mass range of nonzero charge, consistent with the compactness-driven enhancement of the scalar field. In the CFLm model, both the SQS mass at which the scalar charge becomes nonzero and the maximum mass that supports a nonzero scalar charge decrease with $m_s$. These results demonstrate that the scalar charge is potentially a more sensitive diagnostic of scalarization than the mass-radius relation. In particular, the mass-radius sequences in general relativity and scalar-tensor gravity can remain close over a substantial range of masses, while the scalar charge changes from exactly zero in GR to a finite value on the scalarized branch. This distinction is especially important for the weakest scalarization considered here, where the deviations in $M$ and $R$ may be comparatively small. Therefore, an observational inference of a nonzero scalar coupling of a compact object, rather than a large deviation in its mass or radius alone, would provide a direct indication that the object contains a non-trivial scalar field.

An important point is that the scalar field inside an SQS is not expected to be measured directly. Instead, its presence can be inferred through the gravitational effects associated with the asymptotic scalar charge. For an isolated star, the scalar charge determines the scalar component of the exterior field and therefore represents the quantity that connects the internal scalarized configuration to potentially observable phenomena. In a binary system, scalarized components can acquire different scalar charges, and the resulting difference in scalar coupling can generate scalar dipole radiation in addition to the tensor gravitational radiation predicted by GR. Such an additional radiation channel modifies the orbital evolution and the gravitational wave phase. Therefore, an observation of a binary system whose gravitational wave evolution is inconsistent with the GR prediction, together with independently constrained stellar masses and EoSs, could provide evidence for a nonzero scalar charge and hence for a non-trivial scalar field inside the SQS. The present calculations provide the stellar scalar charges required as input for such a future binary analysis.

\begin{figure}[h]
	\subfigure{}{\includegraphics[scale=0.36]{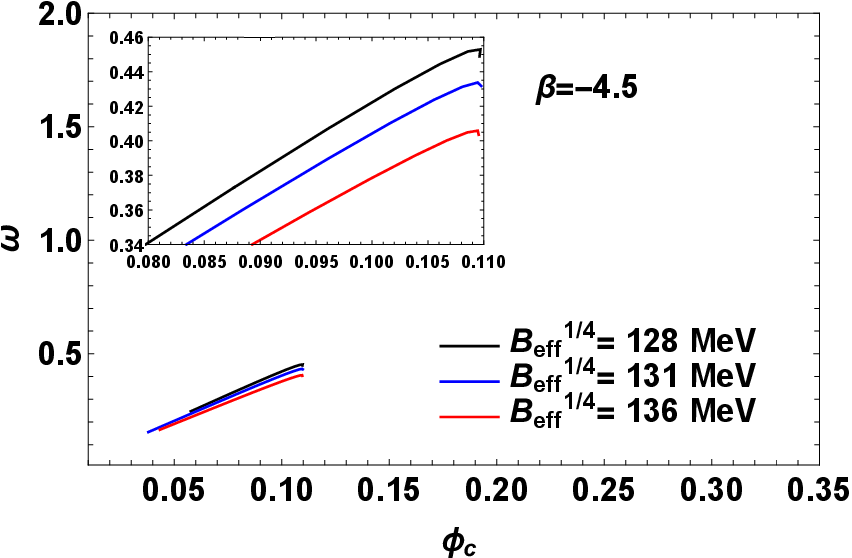}
		}	
\subfigure{}{\includegraphics[scale=0.36]{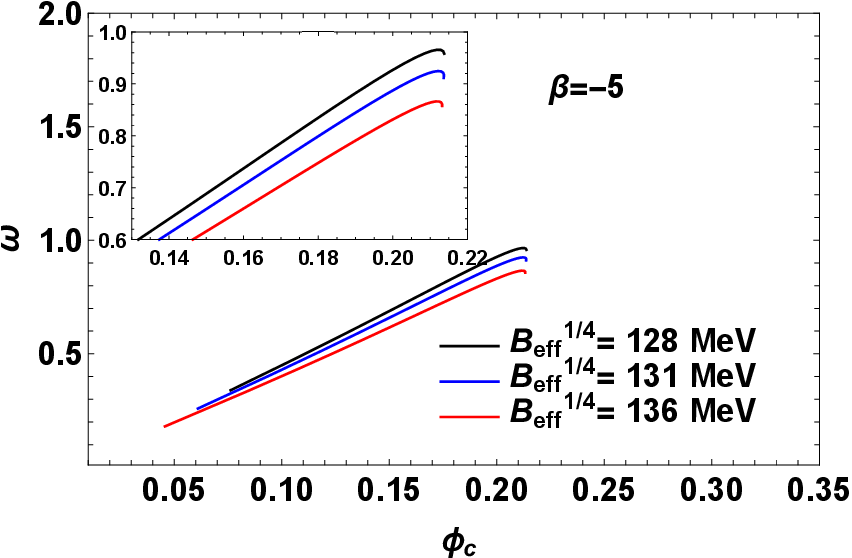}
		}
\subfigure{}{\includegraphics[scale=0.36]{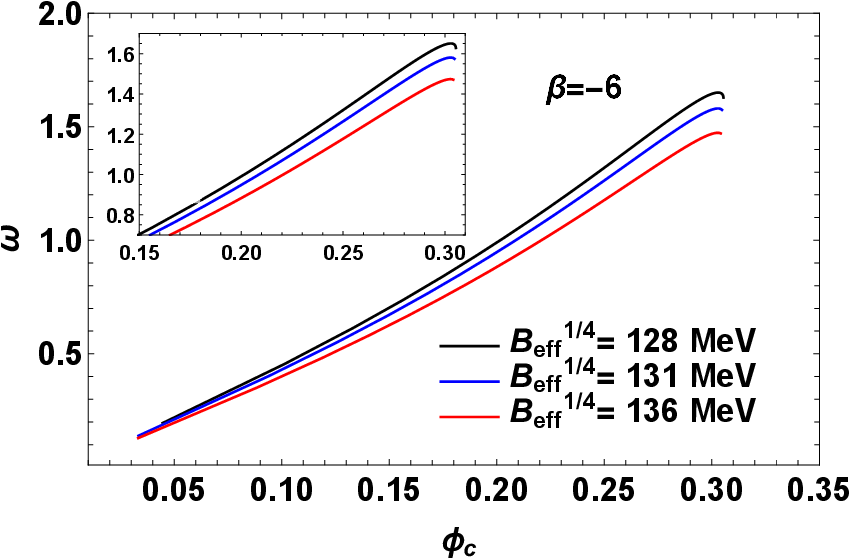}
		}\\
	\subfigure{}{\includegraphics[scale=0.36]{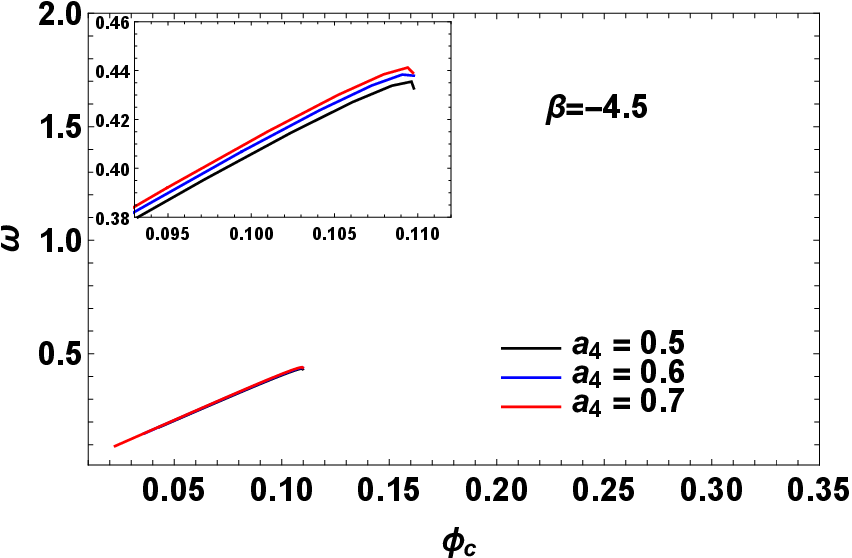}
		}	
\subfigure{}{\includegraphics[scale=0.36]{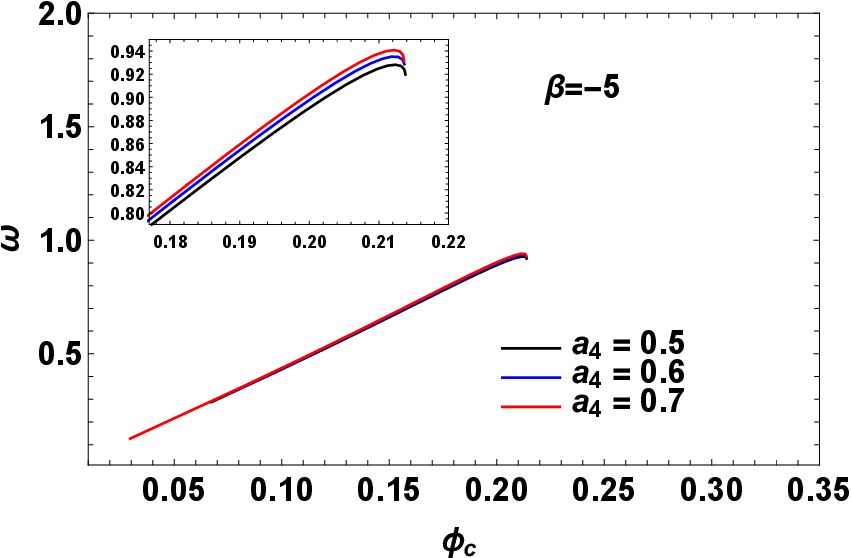}
		}
\subfigure{}{\includegraphics[scale=0.36]{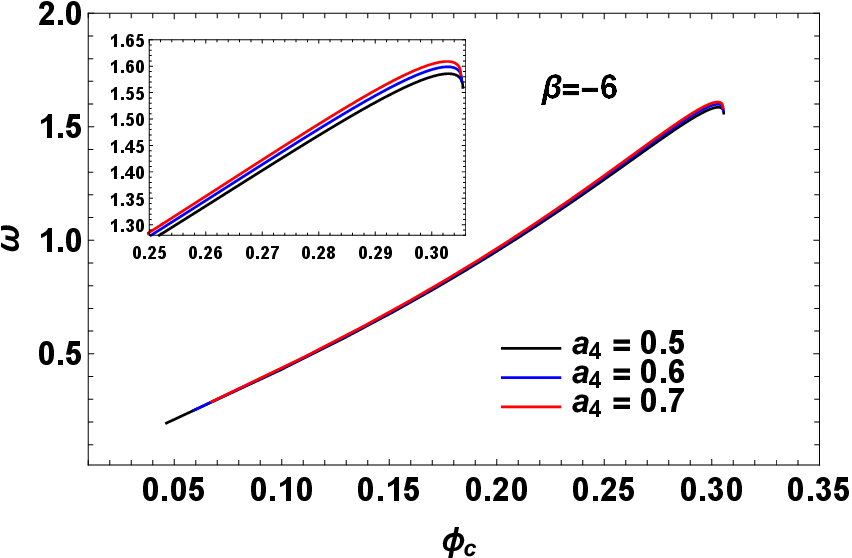}
		}
	\caption{Scalar charge, $\omega$, versus the central scalar field, $\phi_c$, in Normal model with the value of $m_s=100MeV$, Top: considering the value of $a_4=0.58$ and different values of $B_{eff}$, and Bottom: considering the value of $B_{eff}^{1/4}=130.3MeV$ and different values of $a_4$, for three values of the coupling constant $\beta$.}
	\label{omphiNormal}
\end{figure}

\begin{figure}[h]
	\subfigure{}{\includegraphics[scale=0.36]{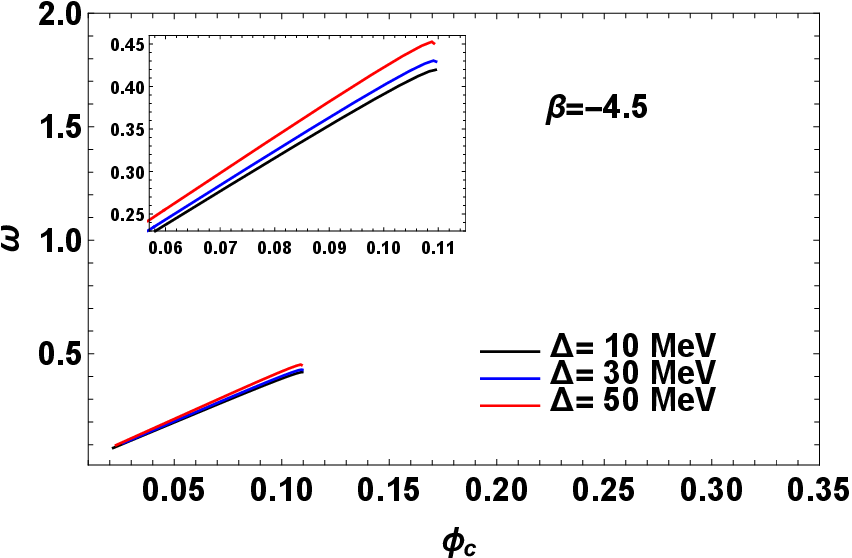}
		}	
	\subfigure{}{\includegraphics[scale=0.36]{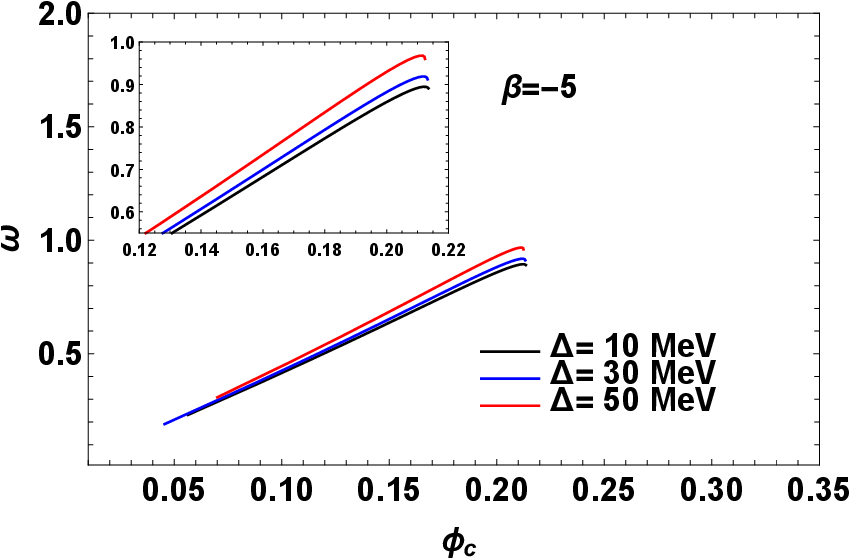}
		}
	\subfigure{}{\includegraphics[scale=0.36]{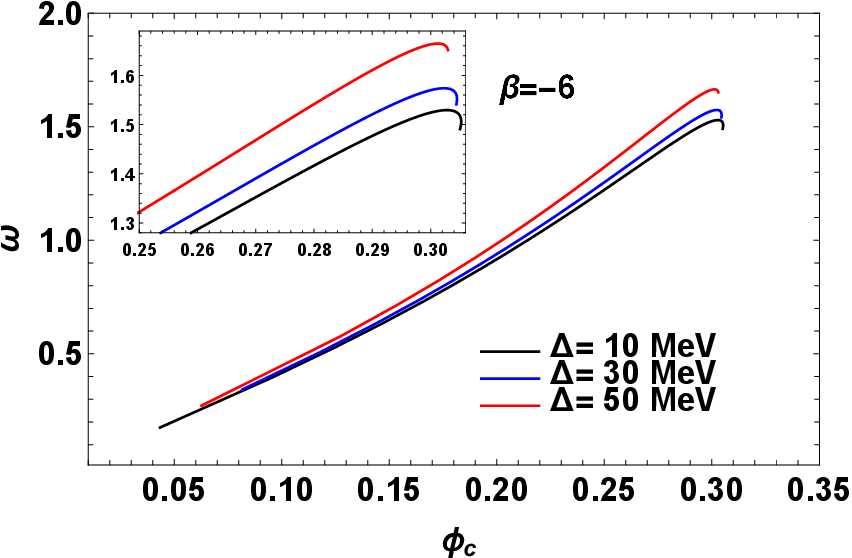}
		}
	\caption{Scalar charge, $\omega$, versus the central scalar field, $\phi_c$, in CFL model with the values of $m_s=100MeV$, $a_4=0.53$, $B_{eff}^{1/4}=133.1MeV$, and different values of $\Delta$ for three values of the coupling constant, $\beta$.}
	\label{omphiCFL}
\end{figure}

\begin{figure}[h]
	\subfigure{}{\includegraphics[scale=0.36]{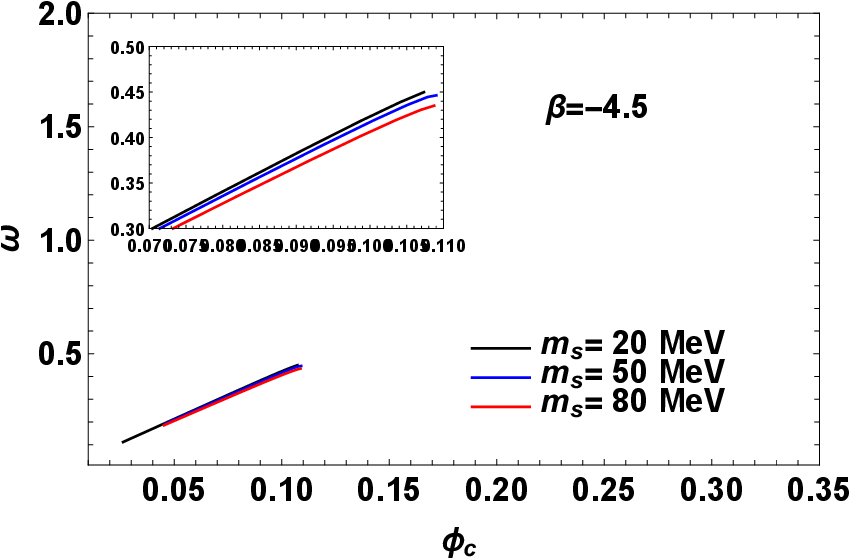}
		}	
	\subfigure{}{\includegraphics[scale=0.36]{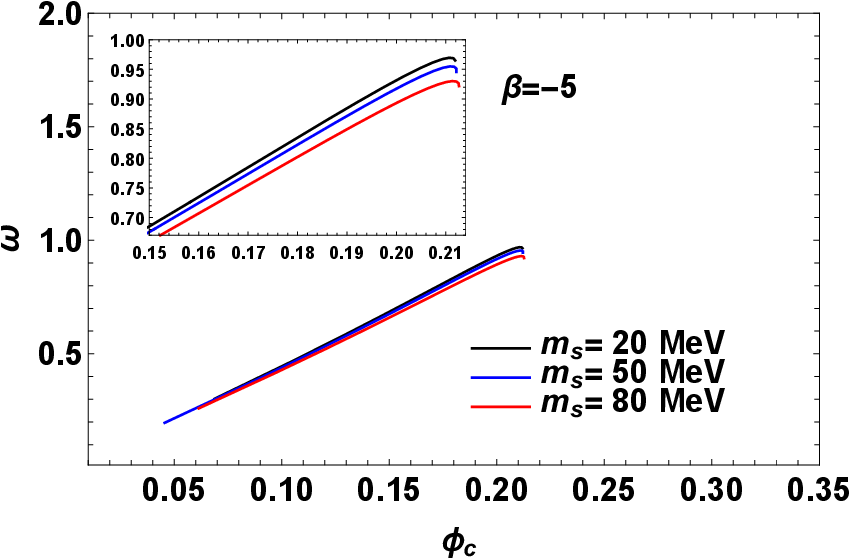}
		}
	\subfigure{}{\includegraphics[scale=0.36]{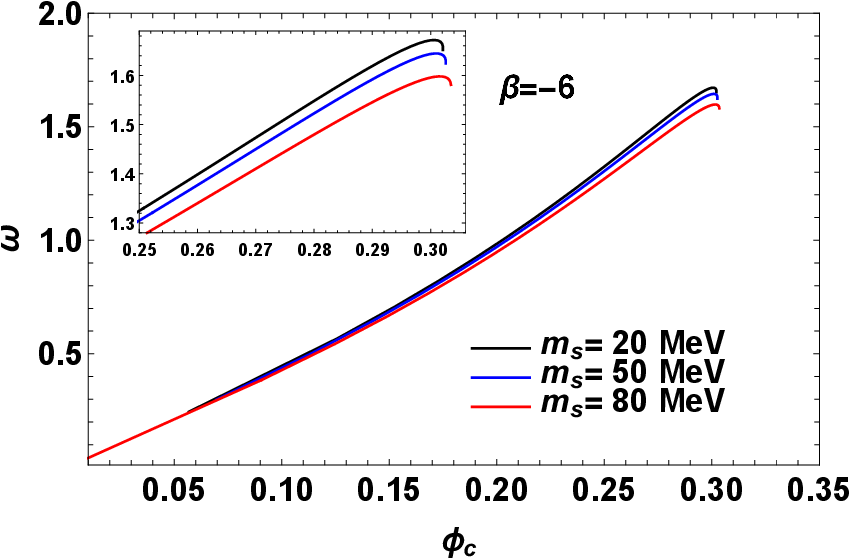}
		}
	\caption{Scalar charge, $\omega$, versus the central scalar field, $\phi_c$, in CFLm model with the values of $a_4=0.53$, $\Delta=33.9MeV$, $B_{eff}^{1/4}=134.4MeV$, and different values of $m_s$ for three values of the coupling constant, $\beta$.}
	\label{omphiCFLm}
\end{figure}

Figures~\ref{omphiNormal}--\ref{omphiCFLm} present the scalar charge
$\omega$ as a function of the central scalar field $\Phi_c$ for the
three families of SQSs. The two quantities characterize
different aspects of the same scalarized configuration: $\Phi_c$
measures the scalar field amplitude at the stellar center, whereas
$\omega$ measures the strength of the scalar field in the asymptotic
exterior. Their relation is obtained by integrating the scalar field
equation from the center to the stellar surface and matching the
interior solution to the exterior solution, as described by
Eq.~(\ref{omM}). Therefore, $\omega$ is determined by $\Phi_c$ only
through the complete stellar structure and is not universally equal or
proportional to it.

For each EoS considered here, $\omega$ increases with
$\Phi_c$ over most of the scalarized branch, reaches a broad maximum,
and then exhibits a mild turnover at large $\Phi_c$. Thus, stronger
internal scalarization generally produces a stronger exterior scalar
charge, although the mapping $\omega(\Phi_c)$ depends on the EoS, the stellar compactness, and the coupling constant $\beta$.
The maximum scalar charge $\omega_{\max}$ depends systematically
on the stiffness of the EoS. At fixed $\Phi_c$, the scalar charge depends on the EoS
because the latter determines the stellar compactness, radial density
profile, and hence the radial evolution of the scalar field between the
center and the surface. The larger $\omega$ obtained for the stiffer
configurations therefore reflects the different stellar structure
associated with the same central scalar field amplitude. This connection is the direct link between the
microphysics of dense quark matter and the dipolar gravitational wave emission
that could be observed from binary systems containing scalarized SQSs. Besides, Figures~\ref{omphiNormal}--\ref{omphiCFLm} denotes that as the coupling strength increases from
$\beta=-4.5$ to $\beta=-6$, $\omega_{\max}$ grows and the corresponding curves extend over a
much wider range.

Figure~\ref{omphiNormal} shows that the
peak scalar charge decreases with increasing $B_{\rm eff}$, while it grows slightly by $a_4$.
This confirms that $B_{\rm eff}$ is a strong lever on the observable scalar
charge, whereas $a_4$ provides only a fine tuning. The reason is that $B_{\rm eff}$
changes the low-density structure and the surface density of the quark star
much more strongly than $a_4$, producing a wider variation in compactness.
Figure~\ref{omphiCFL} indicates that the maximum scalar charge increases
with stiffness (i.e. larger $\Delta$). Moreover, at a fixed central scalar field, a
larger $\Delta$ gives a larger $\omega$. This is the direct imprint of
compactness: a larger pairing gap stiffens the CFL EoS, making
the star more compact, so the same interior scalar field is transmitted to
the asymptotic exterior more efficiently. Fig.~\ref{omphiCFLm} verifies that the peak scalar charge decreases
with increasing $m_s$. The differences are smaller than those produced by $B_{\rm eff}$ or $\Delta$,
indicating that $m_s$ acts as a mild scalarization suppressor, but the
compactness-driven ordering is still visible. These results complete the
picture of the observable scalar charge: for all three families, the internal
scalar field amplitude is controlled by $\beta$, while the exterior scalar
charge also carries a clean and systematic imprint of the EoS stiffness.

An important consequence of these results is that the central scalar
field and scalar charge are correlated but are not interchangeable
diagnostics. The central field provides a measure of the strength of
scalarization in the stellar interior, whereas the scalar charge
provides the corresponding asymptotic quantity relevant for
gravitational interactions outside the star. The EoS
dependence of the $\omega(\Phi_c)$ relation demonstrates that the same
central scalar field amplitude can lead to different scalar charges for
different stellar structures. Hence, the scalar charge contains
additional information about how the interior scalarization is mapped
onto the exterior spacetime.

\section{SUMMARY AND CONCLUDING REMARKS}\label{s}

In this work, we have performed a comprehensive analysis of spontaneous scalarization in strange quark stars within scalar-tensor gravity. Three physically distinct descriptions of self-bound strange quark matter were employed - normal quark matter, colour-flavour-locked (CFL) superfluid matter, and CFL matter with a free strange quark mass (CFLm) - covering a broad range of microphysical parameters: the effective bag constant, $B_{eff}$, the perturbative QCD correction, $a_4$, the pairing gap, $\Delta$, and the strange quark mass, $m_s$. For each equation of state, the generalised Tolman-Oppenheimer-Volkoff equations were solved for different values of the coupling constant and a set of quantities was computed: the mass-central density and mass-radius relations, the central scalar field, the critical density, and the star scalar charge.

The main result of this study is that spontaneous scalarization in strange quark stars is a compactness-driven phenomenon. It is governed by compactness, with the coupling strength, $\beta$, acting as the dominant control and the equation of state parameters providing systematic fine-tuning. At $\beta=-4.5$, scalarization is marginal, producing small deviations in the mass-radius relation, weak central scalar fields, and modest scalar charges. At $\beta=-5$, the scalarization range widens dramatically, the central scalar field grows, the mass-radius branches become clearly distinguishable from their general-relativistic counterparts, and the scalar charge become robust. At $\beta=-6$, scalarization becomes extreme: the mass-radius branches grow enormously with high maximum masses, and the scalar charge curves span nearly the entire stable mass range. Any change in the equation of state that increases the stellar compactness (stiffening the matter by a smaller $B_{eff}$ or $m_s$, or a larger $a_4$ or $\Delta$) deepens the gravitational potential, strengthens the tachyonic instability, and shifts the scalarization interval to lower central densities while raising the maximum scalarized mass. Conversely, softening the equation of state reduces compactness and delays or suppresses scalarization. The bag constant, $B_{eff}$, and the strange quark mass, $m_s$, act as scalarization suppressors; the perturbative parameter, $a_4$, and the pairing gap, $\Delta$, act as scalarization enhancers, with $B_{eff}$ having the largest quantitative influence. The critical density maps the onset of scalarization, showing that stiffer matter shifts the onset to lower central densities, while softer matter shifts it to higher densities. The scalar charge provides a direct observable signature in binary systems: its mass range and amplitude are controlled by $\beta$ and the stiffness of the equation of state, and it can lead to strong dipolar gravitational wave emission.

Our results demonstrate that strange quark stars are viable and potentially observable laboratories for testing scalar-tensor theories of gravity. The combination of mass-radius measurements from X-ray timing, gravitational wave observations of binary mergers, and binary pulsar timing data can probe the scalar charge window and constrain both the gravitational coupling, $\beta$, and the equation of state of cold dense quark matter. The systematic framework established
in this work - linking QCD-scale parameters to macroscopic scalarization observables - can be straightforwardly extended to include rotation, finite temperature, magnetic fields, or dynamical scalarization in mergers, and will be essential for interpreting future multimessenger astrophysical data.

Finally, we emphasize that the present work does not assume that strange quark stars have been observationally established. Rather, they remain one of several viable theoretical possibilities for the composition of compact stars. The primary motivation of this study is to investigate whether spontaneous scalarization introduces additional observable signatures that could help distinguish self-bound strange quark stars from conventional neutron stars when combined with future multimessenger observations. If scalarization-induced effects on quantities such as the scalar charge, mass-radius relation, or gravitational wave emission become observationally accessible, they could provide complementary constraints on both the nature of dense matter and the underlying theory of gravity.

%%%%%%%%%%%%%%%%%%%%%%%%%%%%%%%%%%%%%%%%%%%%%%%%%%%%%%%%%%%%%%%%%%%%%%%%%%%%%%%%%%%%%%%%%%%%%%%%%%%
%%%%%%%%%%%%%%%%%%%%%%%%%%%%%%%%%%%%%%%%%%%%%%%%%%%%%%%%%%%%%%%%%%%%%%%%%%%%%%%%%%%%%%%%%%

\acknowledgements{The authors wish to thank the Shiraz University Research Council.}

\end{document}